\documentclass[
reprint,
superscriptaddress,
amsmath,amssymb,
aps,
pra,
floatfix,
]{revtex4-2}

\usepackage{graphicx}
\usepackage{bm}
\usepackage{float} 
\usepackage[colorlinks=true,allcolors=blue]{hyperref}
\usepackage{xcolor}
\usepackage{multirow}
\usepackage[T1]{fontenc}
\usepackage{newtxtext,newtxmath}
\usepackage{booktabs}

\usepackage{rotating}

\begin{document}

\title{A zero-dead-time strontium lattice clock with a stability at  $10^{-19}$ level}
\makeatletter\let\thetitle\@title\makeatother

\author{Xiao-Yong~Liu$^{\#}$}
\affiliation{Hefei National Research Center for Physical Sciences at the Microscale and School of Physical Sciences, University of Science and Technology of China, Hefei 230026, China}
\affiliation{Shanghai Research Center for Quantum Sciences and CAS Center for Excellence in Quantum Information and Quantum Physics, University of Science and Technology of China, Shanghai 201315, China}

\author{Peng~Liu$^{\#}$}
\affiliation{Shanghai Research Center for Quantum Sciences and CAS Center for Excellence in Quantum Information and Quantum Physics, University of Science and Technology of China, Shanghai 201315, China}
\affiliation{Hefei National Laboratory, University of Science and Technology of China, Hefei 230088, China}

\author{Jie~Li$^{\#}$}
\affiliation{Hefei National Research Center for Physical Sciences at the Microscale and School of Physical Sciences, University of Science and Technology of China, Hefei 230026, China}
\affiliation{Shanghai Research Center for Quantum Sciences and CAS Center for Excellence in Quantum Information and Quantum Physics, University of Science and Technology of China, Shanghai 201315, China}

\author{Yu-Chen~Zhang}
\affiliation{Hefei National Research Center for Physical Sciences at the Microscale and School of Physical Sciences, University of Science and Technology of China, Hefei 230026, China}
\affiliation{Shanghai Research Center for Quantum Sciences and CAS Center for Excellence in Quantum Information and Quantum Physics, University of Science and Technology of China, Shanghai 201315, China}

\author{Yuan-Bo~Wang}
\affiliation{Hefei National Research Center for Physical Sciences at the Microscale and School of Physical Sciences, University of Science and Technology of China, Hefei 230026, China}
\affiliation{Shanghai Research Center for Quantum Sciences and CAS Center for Excellence in Quantum Information and Quantum Physics, University of Science and Technology of China, Shanghai 201315, China}

\author{Zhi-Peng~Jia}
\affiliation{Hefei National Research Center for Physical Sciences at the Microscale and School of Physical Sciences, University of Science and Technology of China, Hefei 230026, China}
\affiliation{Shanghai Research Center for Quantum Sciences and CAS Center for Excellence in Quantum Information and Quantum Physics, University of Science and Technology of China, Shanghai 201315, China}

\author{Xiang~Zhang}
\affiliation{Hefei National Research Center for Physical Sciences at the Microscale and School of Physical Sciences, University of Science and Technology of China, Hefei 230026, China}
\affiliation{Shanghai Research Center for Quantum Sciences and CAS Center for Excellence in Quantum Information and Quantum Physics, University of Science and Technology of China, Shanghai 201315, China}

\author{Xian-Qing~Zhu}
\affiliation{Hefei National Research Center for Physical Sciences at the Microscale and School of Physical Sciences, University of Science and Technology of China, Hefei 230026, China}
\affiliation{Shanghai Research Center for Quantum Sciences and CAS Center for Excellence in Quantum Information and Quantum Physics, University of Science and Technology of China, Shanghai 201315, China}

\author{De-Quan~Kong}
\affiliation{Hefei National Research Center for Physical Sciences at the Microscale and School of Physical Sciences, University of Science and Technology of China, Hefei 230026, China}
\affiliation{Shanghai Research Center for Quantum Sciences and CAS Center for Excellence in Quantum Information and Quantum Physics, University of Science and Technology of China, Shanghai 201315, China}

\author{Wen-Lan~Song}
\affiliation{Hefei National Research Center for Physical Sciences at the Microscale and School of Physical Sciences, University of Science and Technology of China, Hefei 230026, China}
\affiliation{Shanghai Research Center for Quantum Sciences and CAS Center for Excellence in Quantum Information and Quantum Physics, University of Science and Technology of China, Shanghai 201315, China}

\author{Guo-Zhen~Niu}
\affiliation{Shanghai Research Center for Quantum Sciences and CAS Center for Excellence in Quantum Information and Quantum Physics, University of Science and Technology of China, Shanghai 201315, China}
\affiliation{Hefei National Laboratory, University of Science and Technology of China, Hefei 230088, China}

\author{Yu-Meng~Yang}
\affiliation{Shanghai Research Center for Quantum Sciences and CAS Center for Excellence in Quantum Information and Quantum Physics, University of Science and Technology of China, Shanghai 201315, China}
\affiliation{Hefei National Laboratory, University of Science and Technology of China, Hefei 230088, China}

\author{Pei-Jun Feng}
\affiliation{Shanghai Research Center for Quantum Sciences and CAS Center for Excellence in Quantum Information and Quantum Physics, University of Science and Technology of China, Shanghai 201315, China}
\affiliation{Hefei National Laboratory, University of Science and Technology of China, Hefei 230088, China}

\author{Xiang-Pei~Liu}
\affiliation{Hefei National Research Center for Physical Sciences at the Microscale and School of Physical Sciences, University of Science and Technology of China, Hefei 230026, China}
\affiliation{Shanghai Research Center for Quantum Sciences and CAS Center for Excellence in Quantum Information and Quantum Physics, University of Science and Technology of China, Shanghai 201315, China}
\affiliation{Hefei National Laboratory, University of Science and Technology of China, Hefei 230088, China}

\author{Xing-Yang~Cui}
\affiliation{Shanghai Research Center for Quantum Sciences and CAS Center for Excellence in Quantum Information and Quantum Physics, University of Science and Technology of China, Shanghai 201315, China}
\affiliation{Hefei National Laboratory, University of Science and Technology of China, Hefei 230088, China}

\author{Ping~Xu}
\affiliation{Hefei National Research Center for Physical Sciences at the Microscale and School of Physical Sciences, University of Science and Technology of China, Hefei 230026, China}
\affiliation{Shanghai Research Center for Quantum Sciences and CAS Center for Excellence in Quantum Information and Quantum Physics, University of Science and Technology of China, Shanghai 201315, China}
\affiliation{Hefei National Laboratory, University of Science and Technology of China, Hefei 230088, China}

\author{Xiao~Jiang}
\affiliation{Hefei National Research Center for Physical Sciences at the Microscale and School of Physical Sciences, University of Science and Technology of China, Hefei 230026, China}
\affiliation{Shanghai Research Center for Quantum Sciences and CAS Center for Excellence in Quantum Information and Quantum Physics, University of Science and Technology of China, Shanghai 201315, China}
\affiliation{Hefei National Laboratory, University of Science and Technology of China, Hefei 230088, China}

\author{Juan~{Yin}}
\affiliation{Hefei National Research Center for Physical Sciences at the Microscale and School of Physical Sciences, University of Science and Technology of China, Hefei 230026, China}
\affiliation{Shanghai Research Center for Quantum Sciences and CAS Center for Excellence in Quantum Information and Quantum Physics, University of Science and Technology of China, Shanghai 201315, China}
\affiliation{Hefei National Laboratory, University of Science and Technology of China, Hefei 230088, China}

\author{Sheng-Kai~{Liao}}
\affiliation{Hefei National Research Center for Physical Sciences at the Microscale and School of Physical Sciences, University of Science and Technology of China, Hefei 230026, China}
\affiliation{Shanghai Research Center for Quantum Sciences and CAS Center for Excellence in Quantum Information and Quantum Physics, University of Science and Technology of China, Shanghai 201315, China}
\affiliation{Hefei National Laboratory, University of Science and Technology of China, Hefei 230088, China}

\author{Cheng-Zhi~{Peng}}
\affiliation{Hefei National Research Center for Physical Sciences at the Microscale and School of Physical Sciences, University of Science and Technology of China, Hefei 230026, China}
\affiliation{Shanghai Research Center for Quantum Sciences and CAS Center for Excellence in Quantum Information and Quantum Physics, University of Science and Technology of China, Shanghai 201315, China}
\affiliation{Hefei National Laboratory, University of Science and Technology of China, Hefei 230088, China}

\author{Han-Ning~{Dai}}
\email[]{daihan@ustc.edu.cn}

\affiliation{Hefei National Research Center for Physical Sciences at the Microscale and School of Physical Sciences, University of Science and Technology of China, Hefei 230026, China}
\affiliation{Shanghai Research Center for Quantum Sciences and CAS Center for Excellence in Quantum Information and Quantum Physics, University of Science and Technology of China, Shanghai 201315, China}
\affiliation{Hefei National Laboratory, University of Science and Technology of China, Hefei 230088, China}

\author{Yu-Ao~{Chen}}
\email[]{yuaochen@ustc.edu.cn}
\affiliation{Hefei National Research Center for Physical Sciences at the Microscale and School of Physical Sciences, University of Science and Technology of China, Hefei 230026, China}
\affiliation{Shanghai Research Center for Quantum Sciences and CAS Center for Excellence in Quantum Information and Quantum Physics, University of Science and Technology of China, Shanghai 201315, China}
\affiliation{Hefei National Laboratory, University of Science and Technology of China, Hefei 230088, China}
\affiliation{New Cornerstone Science Laboratory, School of Emergent Technology, University of Science and Technology of China, Hefei 230026, China}

\author{Jian-Wei~{Pan}}
\email[]{pan@ustc.edu.cn}
\affiliation{Hefei National Research Center for Physical Sciences at the Microscale and School of Physical Sciences, University of Science and Technology of China, Hefei 230026, China}
\affiliation{Shanghai Research Center for Quantum Sciences and CAS Center for Excellence in Quantum Information and Quantum Physics, University of Science and Technology of China, Shanghai 201315, China}
\affiliation{Hefei National Laboratory, University of Science and Technology of China, Hefei 230088, China}

\date{\today}

\begin{abstract}
\noindent Optical atomic clocks play a crucial role in fundamental physics, relativistic geodesy, and the future redefinition of the SI second. Standard operation relies on cyclic interrogation sequences, which alternate between atomic interrogation and dead time used for state preparation and readout. This approach introduces the Dick effect, where laser frequency noise aliases onto the atomic transition frequency. Although reducing laser noise improves clock stability, the Dick effect remains a key limitation. In this work, we demonstrate a zero-dead-time optical clock based on two interleaved ensembles of cold \(^{87}\text{Sr}\) atoms. Our system significantly suppresses this noise and achieves a fractional frequency instability at the \(10^{-19}\) level between 10,000 and 20,000 seconds over repeated measurements, with a best value of \(2.9 \times 10^{-19}\) at \(\tau = 20,000\) seconds. The estimated long-term stability based on the combined data of these measurements reaches \(2.5 \times 10^{-19}\) at one day. 
These results represent a more than ninefold improvement over a conventional single-ensemble clock, highlighting its potential for next-generation timekeeping applications.
\end{abstract}

\maketitle

\section{Introduction}\label{sec1}

Optical atomic clocks are leading candidates for next-generation time-frequency standards due to their exceptional precision in quantum metrology~\cite{ludlow2015optical}.
Advances in laser cooling and trapping, quantum state control, and ultra-stable lasers have enabled these clocks to achieve fractional stability and uncertainty below $10^{-18}$~\cite{mcgrew2018,aeppli2024,oelker2019,bothwell2019,marshall2025}. Optical clocks have now been extensively developed in various neutral atom~\cite{ludlow2015optical,mcgrew2018,aeppli2024,takamoto2020} and trapped-ion~\cite{marshall2025,Zhang2023,Tofful2024,Hausser2025,Zhang2025} systems. Notably, high-precision clock comparisons have been demonstrated in gravitational redshift measurements~\cite{mcgrew2018,takamoto2020}. In addition, new technologies have been proposed, expected to further enhance the performance limits of optical clocks, such as zero-dead-time (ZDT) architecture~\cite{schioppo2017ultrastable,kim2023} and entangled optical clocks~\cite{nichol2022}.
These breakthroughs underpin the impending redefinition of the SI second~\cite{dimarcq2024roadmap,lodewyck2019definition,riehle2015towards} and support broad-ranging applications in fundamental physics, including tests of variations in fundamental constants~\cite{huntemann2014improved}, precision tests of general relativity~\cite{bothwell2022resolving,takamoto2020}, and the detection of gravitational waves and dark matter candidates~\cite{naroznik2023,abbott2020,su2018,kolkowitz2016,filzinger2025,arvanitaki2015}, extending far beyond metrology into the frontier of science.

The stability of optical clocks is fundamentally limited by the quantum projection noise (QPN) and technically constrained by noise from the optical local oscillator (OLO). 
The QPN scales as $1/\sqrt{N}$ with the number of atoms $N$, approaching a theoretical limit of less than $5.0 \times 10^{-17}/\sqrt{\tau}$ for strontium clocks with $N \sim 10^3$~\cite{itano1993}. 
While quantum entanglement could potentially surpass this limit~\cite{robi2024,cao2024}, current performance remains typically limited to $\sim 10^{-16}/\sqrt{\tau}$ due to the Dick effect, aliasing of OLO noise into atomic signals via dead time during state preparation and readout~\cite{dick1989local}. 
Several strategies have been developed to suppress the Dick effect:  enhancing ultra-stable lasers, which requires substantial improvements in optical cavity stability ~\cite{parke2025,kedar2023,yu2023,robinson2019,oelker2019}; employing longitudinal Ramsey spectroscopy in a moving optical lattice, which necessitates precise stabilization of laser and magnetic fields~\cite{katori2021}; or ZDT architectures using alternately interrogated atomic ensembles, which demands two independent interrogation channels~\cite{kim2023,schioppo2017ultrastable,biedermann2013}. These approaches represent a significant step toward realizing clocks beyond the Dick effect limit.

While the ZDT architectures are well-established in microwave atomic clocks~\cite{biedermann2013,lin2017,cheng2018}, their implementation in optical clocks remains rare. Schioppo et al.~\cite{schioppo2017ultrastable} demonstrated this approach in a $^{171}$Yb optical lattice clock with two cold-atom ensembles, estimating a potential instability of \hbox{$\sim 6 \times  10^{-17}/\sqrt{\tau}$} with a single OLO in a synchronized fashion. The method's inherent common-mode noise rejection~\cite{takamoto2011} prevents full characterization of the OLO's stability.
Subsequent work~\cite{kim2023} improved the performance of an Al$^+$ ion clock by an order of magnitude via a phase-coherent optical link to the Yb lattice clock. 
However, direct, independent comparison of clocks operating under ZDT conditions to sub-$10^{-18}$ stability has not yet been experimentally demonstrated, highlighting a critical gap in the validation of this technique for optical frequency standards to date.

\begin{figure*}[t]
\centering
\includegraphics[width=0.85\linewidth]{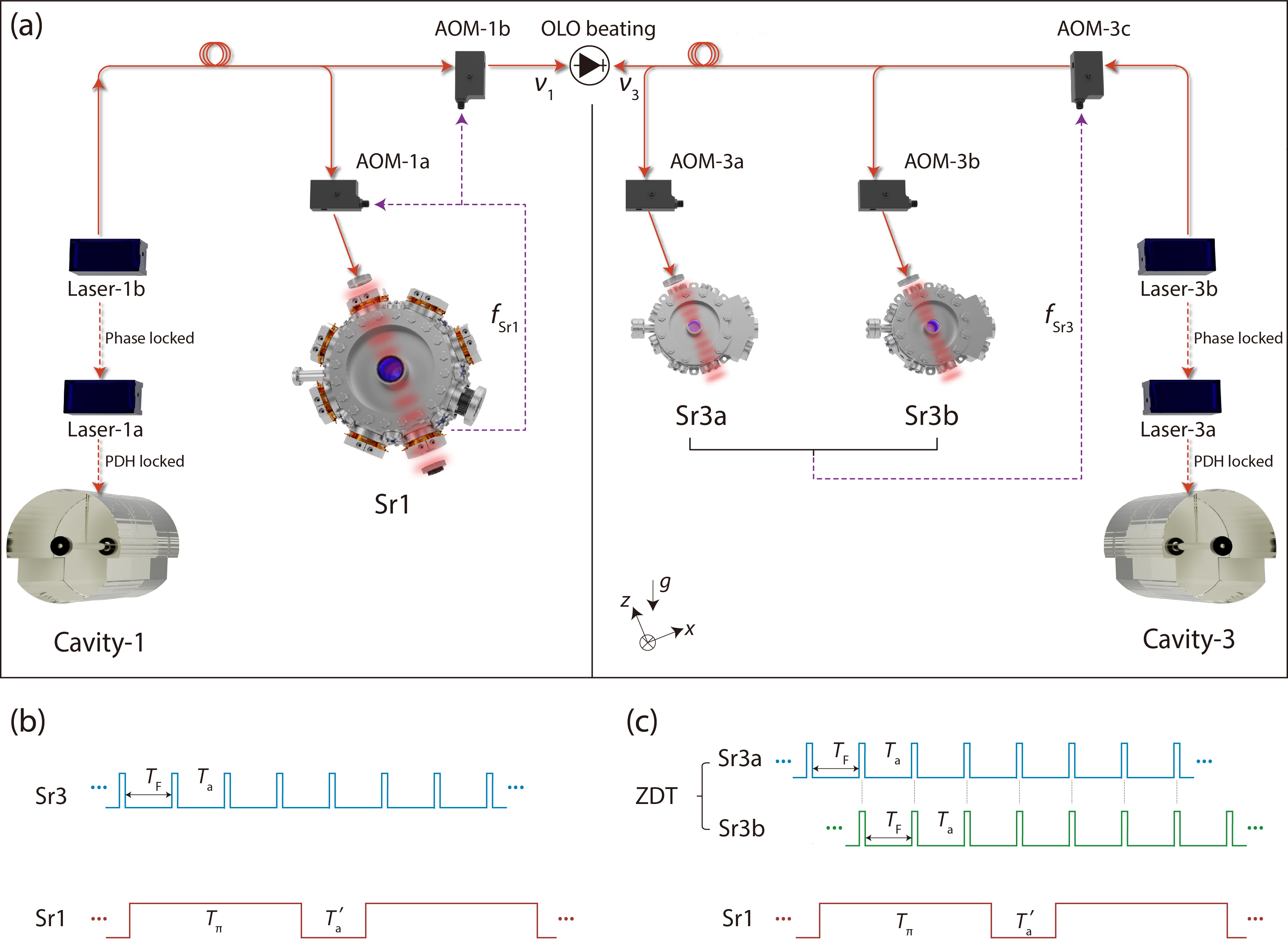}
\caption{
Simplified experimental scheme. (a) Clock comparison layout. The lasers, Laser-1a (1397 nm) and Laser3 (698 nm), are frequency-stabilized to the Cavity-1 and Cavity-3 ULE cavities respectively through Pound-Drever-Hall (PDH) technique. The 698 nm light is then delivered to the optical lattice clock via 70 m long fibers with fiber noise cancellation (FNC). $^{87}$Sr atoms are cooled and loaded into one-dimensional optical lattices within three clocks: Sr3a, Sr3b, and Sr1. The lattice orientation here has a 22.5 angle relative to the vertical direction. The clock lasers (yellow arrows) undergo frequency correction via acousto-optic modulators (AOMs) shown in the figure. The feedback signals applied to their frequencies are indicated by dashed lines. For the ZDT clock and the Sr1 clock, the frequency-corrected laser outputs are denoted $\nu_1$ and $\nu_3$, respectively. These are beat against each other at a photodetector (PD) to generate the beat signal $f_\mathrm{beat}$. This signal is recorded and used to derive clock comparision stability. (b) and (c) show the timing sequences for comparisons with the Sr1 clock during single-clock operation and ZDT operation, respectively. $T_\mathrm{F}$=500ms is Ramsey free evolution time. $T_\mathrm{a}$ and $T^\prime_\mathrm{a}$ are the time for atom loading, state preparation and readout, which are 500ms and 520ms. $T_\pi$=1400ms is the Rabi $\pi$ pulse. The dashed lines in (c) indicate that the $\pi/2$ pulses(25ms) of Sr3a and Sr3b are strictly aligned. 
}
\label{fig1}
\end{figure*}

In this work, we perform independent comparisons between two $^{87}$Sr optical lattice clocks: a high-performance two-ensemble clock working with ZDT scheme and a stability-optimized reference clock. 
Both systems achieve a stability level of $10^{-19}$ while operating with autonomous OLOs. 
This result confirms the ZDT architecture's superiority in long-term stability, while also demonstrating its ability to mitigate stringent short-term oscillator stability requirements. This approach thus holds significant promise for practical, transportable optical clock systems, extending the applicability of ZDT architectures.

\begin{figure*}[t]
	\centering
	\includegraphics[width=0.6\linewidth]{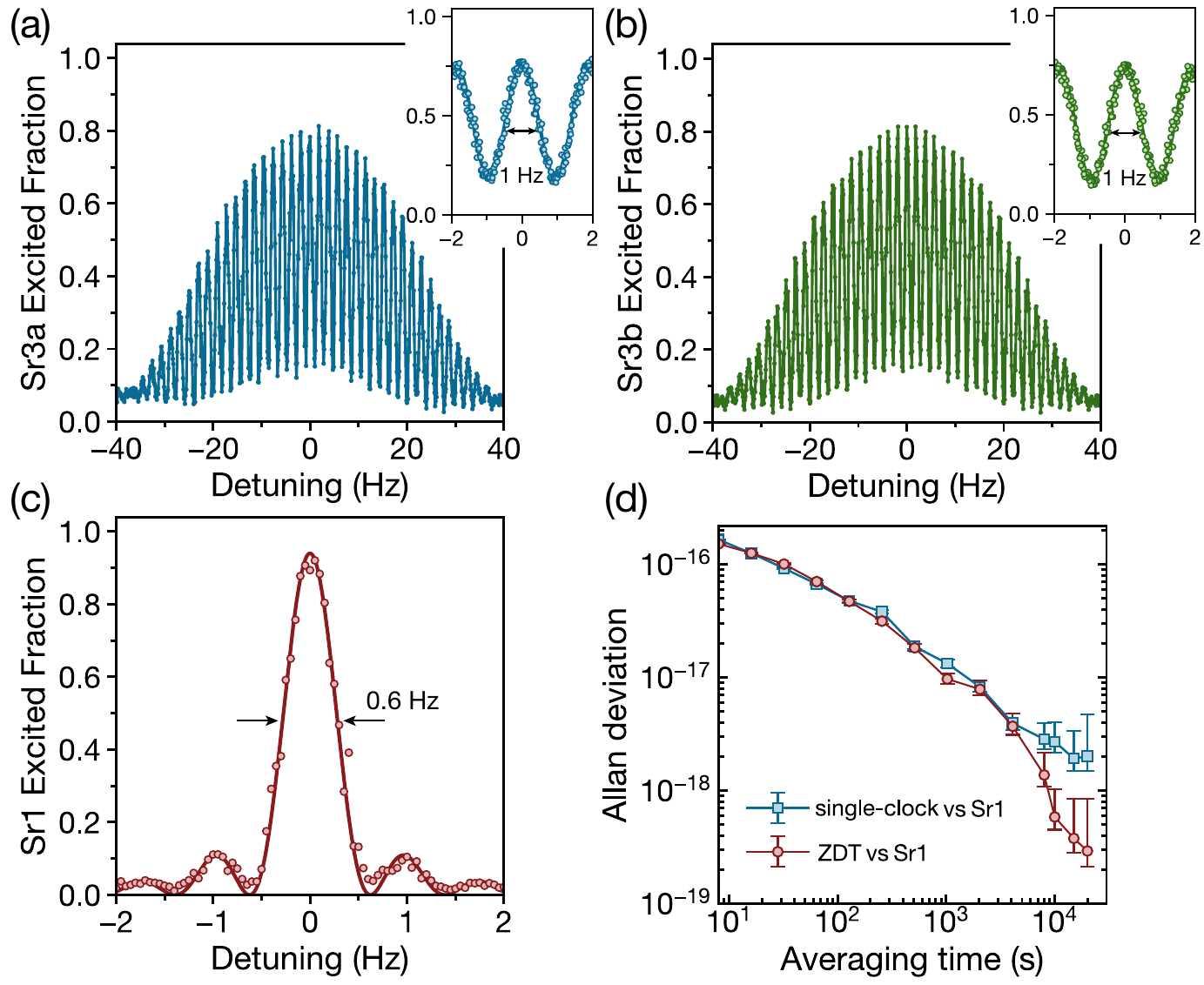}
	\caption{Characterization of the clocks. (a) and (b) are Ramsey spectra of Sr3a and Sr3b for atomic clock transition $|{}^1\mathrm{S}_0, m_F = \pm 9/2\rangle \rightarrow |{}^3\mathrm{P}_0, m_F = \pm 9/2\rangle$. Each data point represents the average of 16 repeated measurements. The linewidth of the central peak is 1 Hz. (c) Rabi spectrum of Sr1 for magnetic-insensitive clock transition $|^{1}\mathrm{S}_0, m_F = \pm 5/2\rangle \rightarrow |^{3}\mathrm{P}_0, m_F = \pm 3/2\rangle$ with 0.6 Hz linewidth. Each data point represents the average of 25 repeated measurements. (d) The stabilities of the clock comparisions, demonstrating the performance enhancement of ZDT (red) relative to the stability in the comparison between single-clock Sr3a and Sr1 (blue). The Allan deviation of the fractional measurement stability is reported as $(\nu_1-\nu_3)/\sqrt{2}\nu_{\mathrm{Sr}}$. 
	}
	\label{fig2}
\end{figure*}

\section{Experimental Setup}\label{sec2}

The experimental setup, illustrated in Fig.~\ref{fig1}(a), consists of two optical lattice clocks, designated as Sr1 and Sr3. These two clocks are located in two different laboratories on separate floors with a height difference of about 4.3 m. Each clock uses its own lasers including atomic cooling and clock lasers and operates with its own timing control system, thus they are fully independent.

The Sr1 atomic system has been thoroughly characterized in previous work~\cite{li2024strontium,Li2023,yu2025}. We implement following upgrades to the system~\cite{jia2025,supp}: an externally mounted cavity outside the vacuum apparatus with a larger lattice beam waist radius of 155~$\mu$m, yielding a $60\times$ intensity enhancement; the use of magnetically insensitive clock transitions for interrogation; and enhanced in-lattice cooling protocols. 
The Sr1 clock operates with $\sim$ 400 atoms at a trap depth of 100 $E_r$, exhibiting axial and radial temperatures of 0.69(3) $\mu$K and 0.52(1) $\mu$K, respectively. Laser-1a at 1397 nm is frequency-stabilized to Cavity-1 via the Pound-Drever-Hall (PDH) technique, with its wavelength converted to 698 nm using a periodically poled lithium niobate (PPLN) crystal. Cavity-1 is a 30-cm-long room-temperature ultra-low expansion (ULE) cavity with crystalline coatings, exhibiting a fractional frequency instability below $2.0\!\times\! 10^{-16}$ between 1 and 50 s~\cite{zhu2024}.  Laser-1b at 698 nm achieves frequency-stabilization by phase-locking to the converted 698 nm source. Its output passes through +80 MHz acousto-optic modulators (AOMs) to drive the clock transitions. During the close-loop operation of the clock, frequency corrections ($f_\mathrm{Sr1}$) are fed back to both AOM-1a and AOM-1b. This stabilized frequency is denoted as $ \nu_1 $ in Fig.~\ref{fig1}(a). A Rabi spectroscopy on the magnetically insensitive transition $|^{1}\mathrm{S}_0, m_F = \pm 5/2\rangle \rightarrow |^{3}\mathrm{P}_0, m_F = \pm 3/2\rangle$ is performed under a 560 mG bias field by a 1400 ms $\pi$-pulse, which exhibits a linewidth of 0.6 Hz, as shown in Fig.~\ref{fig2}.

\begin{figure*}[t]
	\centering
	\includegraphics[width=0.6\linewidth]{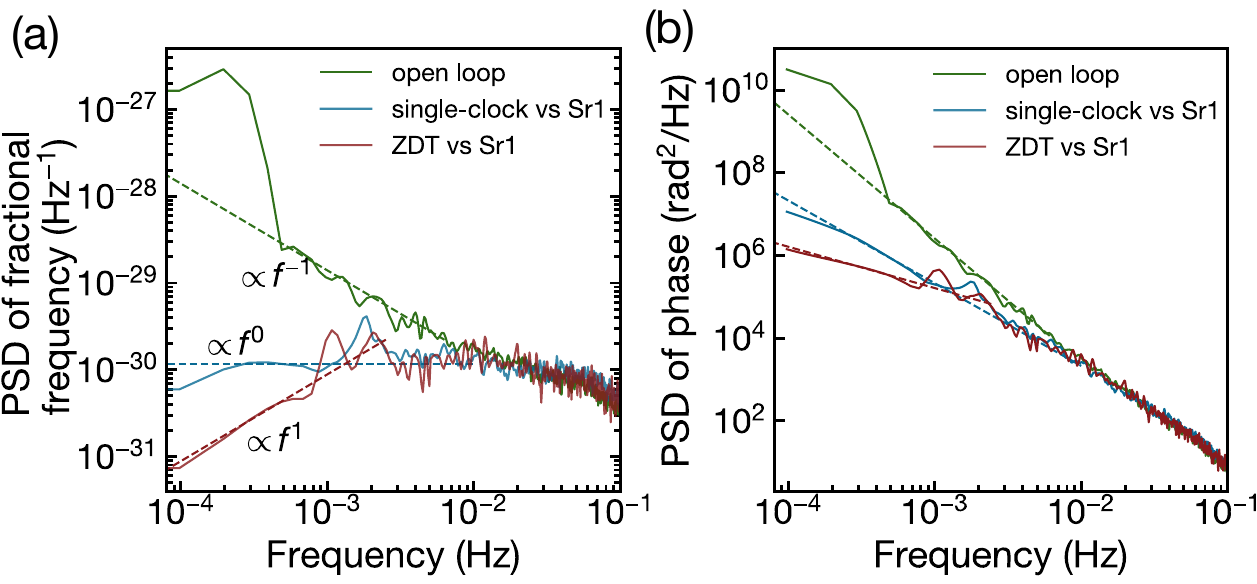}
	\caption{Power spectral densities of beat frequency noise and phase noise. The solid lines of different colors show the noise spectra in different conditions shown in the legend. The dashed lines represent the corresponding noise model fits for each operating condition. The frequency PSD trends follow $f^{-1}$, $f^{0}$, and $f^{1}$ scalings, while the phase PSD trends follow $f^{-3}$, $f^{-2}$, and $f^{-1}$ scalings.
	}
	\label{fig3}
\end{figure*}

The Sr3 atomic system, located in another laboratory, incorporates two identical, compact subsystems (Sr3a and Sr3b), each comprising an atom source and a science chamber. Each subsystem has a total volume of $\sim$ 82~L including optics. These two subsystems are spaced 1 m apart on the optical platform. 
The atomic beam emitted from the dispenser is slowed by a 2D magneto-optical trap (MOT) and injected into the science chamber by a push laser beam. Within the science chamber, 
the $^{87}$Sr atoms are sequentially cooled: first to a few millikelvin via the 461~nm ${}^1\mathrm{S}_0 \rightarrow {}^1\mathrm{P}_1$ transition MOT, and subsequently to a few microkelvin via the 689~nm ${}^1\mathrm{S}_0 \rightarrow {}^3\mathrm{P}_1$ transition narrow-linewidth MOT. Approximately 400 atoms at $\sim$ 2~$\mu$K are loaded into a one-dimensional magic-wavelength optical lattice with a trap depth of 80~$E_r$ and a waist radius of 48~$\mu$m.
These atoms are spin-polarized into the $|{}^1\mathrm{S}_0, m_F = \pm 9/2\rangle$ state using a 500~mG bias field. The clock laser is referenced to Cavity-3, a 30-cm-long room-temperature ULE cavity exhibiting a one-second instability of $2.0 \times 10^{-16}$~\cite{supp}, employing PDH locking and phase locking identical to those used in Sr1, except for Laser-3a operating at 698~nm. During the close-loop operation of Sr3, the OLO frequency corrections ($f_\mathrm{Sr3}$) are applied to AOM-3c, and the resulting output frequency is denoted $\nu_3$ in Fig.~\ref{fig1}(a).
Ramsey spectroscopy is performed to probe the $|{}^1\mathrm{S}_0, m_F = \pm 9/2\rangle \rightarrow |{}^3\mathrm{P}_0, m_F = \pm 9/2\rangle$ clock transition. The Ramsey spectra of Sr3 are shown in Fig.~\ref{fig2}(a-b), acquired with 25-ms $\pi/2$-pulses and a 500-ms free evolution period, resulting in central peaks with a linewidth of 1~Hz.

\section{Clock stability}\label{sec3}

We assess the clock stability by direct comparison of the two independent optical lattice clocks through the OLO beat note measurement, as shown in Fig.~\ref{fig1}(a). The frequency-corrected OLOs outputs are combined at a photodetector, generating a beat signal $f_\mathrm{beat}(t) = \nu_{1}(t)-\nu_{2}(t)$, sampled at a rate of 100~Hz. The Allan deviation of the fractional measurement instability is reported as $f_\mathrm{beat}(t)/\sqrt{2}\nu_{\mathrm{Sr}}$, where $\nu_{\mathrm{Sr}}$ is the $^{87}$Sr clock transition frequency. 

We conduct clock comparisons between Sr1 and Sr3 under two distinct operating modes to characterize the stability. These modes correspond to Sr3 operating in  single-clock mode, where only Sr3a(b) is active, and ZDT mode, where both Sr3a and Sr3b are active. Throughout the measurements, long-term temperature fluctuations are stabilized within 20~mK. Concurrently, real-time compensation of the Sr1 blackbody radiation (BBR) shifts is applied to AOM-1b. These conditions minimize the systematic drifts due to BBR shifts. The configuration with unsynchronized timing sequence and independent OLOs prevent common-mode noise rejection, rendering the comparisons sensitive to all sources of instability including the Dick effect, QPN, fluctuating systematic offsets and other sources of technical noise.

In single-clock mode, the frequency-correction signal $f_\mathrm{Sr3}$ applied to AOM-3c is generated solely by Sr3a.
The timing sequence comprises a 500 ms dead time ($T_\mathrm{a}$), followed by two $\pi/2$-pulses of 25~ms duration, separated by a 500~ms Ramsey free evolution period ($T_\mathrm{F}$), enabling timing-matched alternating interrogation, as shown in Fig.~\ref{fig1}(b). The Sr1 clock operates with a Rabi interrogation time of 1400 ms ($T_\mathrm{\pi}$) and a dead time of 520 ms ($T^\prime_\mathrm{a}$). As shown in Fig.~\ref{fig2}(d), the comparison measurement attains an instability of $2.8\times10^{-16}/\sqrt{\tau}$ for averaging time from 4,000 to 20,000 s, in agreement with the anticipated limit in Table~\ref{table1} accounting for the contributions of both Sr1 and Sr3a. For averaging times below 4,000 s, the Allan deviation data points do not strictly adhere to the $1/\sqrt{\tau}$ scaling trend, indicating the presence of non-common-mode noise between the two OLOs. The final data point at 20,000~s corresponds to a stability of $2.0\!\times\!10^{-18}$. Since the instability of the comparison is primarily dominated by Sr3a(b), the instability of Sr3a(b) can be projected to be approximately $2.8\!\times\!10^{-18}$ at 20,000 seconds.

When Sr3 operates in ZDT mode, the two subsystems perform anti-synchronous interrogation, as shown in Fig.~\ref{fig1}(c), and their combined atomic responses generating a shared OLO correction signal. The overlap of Ramsey $\pi/2$-pulses is synchronized to within 100~ns. 
This scheme significantly suppresses the Dick effect, which is quantitatively described by the sensitivity function. The composite sensitivity function remains approximately unity across all times, with minor deviations only during the Ramsey pulses, enabling continuous monitoring of the OLO frequency~\cite{schioppo2017ultrastable}. 
The Sr1 clock maintains the same operational conditions as in the previous case. Such a comparison achieves $2.9\times10^{-19}$ stability at 20,000~s, which is the best result among multiple measurements with stability at the $10^{-19}$ level~\cite{supp}. It represents a more than ninefold improvement over the single-clock mode, which is dominated by the Dick effect.
 For averaging time beyond 4,000 s, the Allan deviation closely follows a $1/\tau$ scaling trend.

To elucidate the underlying mechanisms governing long-term frequency stability across different operational modes, as illustrated in Fig.~\ref{fig2}(d), we perform spectral noise modeling on the measured $f_\mathrm{beat}(t)$ data, from which we extract key spectral signatures that reveal the dominant noise mechanisms limiting long-term stability, as summarized in Fig.~\ref{fig3}.
In open-loop operation, where the two OLOs are not actively stabilized to the atomic transitions, the power spectral density (PSD) of $f_\mathrm{beat}(t)/\nu_\mathrm{Sr}$ exhibits $f^{-1}$ characteristics. This signature is primarily attributed to thermal noise of the ULE cavities and residual environmental temperature drift. In the single-clock operation, the observed trend exhibits white frequency noise characteristics $f^0$, primarily originating from the Dick effect and the corresponding Allan deviation demonstrates a $1/\sqrt{\tau}$ dependence. In the ZDT operation, the observed $f^1$ scaling yields an stability trend characterized by $1/\tau$ dependence, reflecting the suppression of low-frequency noise. The phase PSD trends are fitted to $f^{-3}$, $f^{-2}$ and $f^{-1}$ scalings for the three cases in Fig~\ref{fig3}(b), validating the consistency of this analysis.

\section{Expectation of Long-term stability}\label{sec4}

\begin{table}[b]
  \centering
  \caption{\centering Instability budget for clocks ($\times10^{-17}/\sqrt{\tau}$).}
  \begin{tabular}{lcccccc}
  \toprule[1pt]
    & & Sr1 & & Sr3a(b) & &ZDT \\
    \midrule[0.5pt]
    $\mathrm{QPN}$ & &$2.0$ & &$3.4$ & & $2.4$ \\
    $\mathrm{Detection}$ & & $2.3$ & & $3.8$ & & $2.7$\\
    $\mathrm{FNC}$ & & $1.0$ & & $1.0$ & & $1.0$\\  
    $\mathrm{Dick}$ & & $9.0$ & & $38.5$ & & $<0.3$\\
    \midrule[0.5pt]
    $\mathrm{\mathbf{Total}}$ & & $\mathbf{9.7}$ & & $\mathbf{38.8}$ & & $\mathbf{4.1}$ \\
  \bottomrule[1pt]
  \end{tabular}
  \label{table1}
\end{table}

% The long-term frequency stability of optical clocks is constrained by fundamental noise sources and is also influenced by variations in systematic shifts over extended periods. We estimate the stability budget of each clock (Table~\ref{table1}), including contributions from QPN, atom detection noise, FNC and Dick effect. 

\begin{figure}[t]
\centering
\includegraphics[width= 0.85\linewidth]{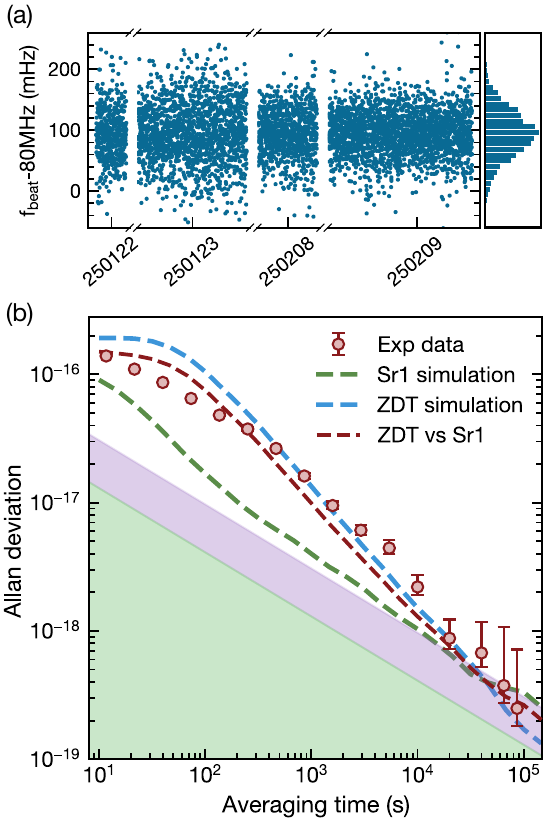}
\caption{Characterization of reproducibility. (a) presents four datasets from comparison experiments conducted at different times, with durations of 7.4 h, 22.6 h, 14.5 h, and 35.4 h, respectively. Data collected during the measurement campaign are binned into 1 minute-long intervals, and the mean frequency for each filled bin is shown as a blue data point. (b) The green and blue lines represent the simulated stability of the Sr1 and ZDT, denoted as $\sigma_1$ and $\sigma_2$, respectively. Their fits are $1.0\times10^{-16}/\sqrt{\tau}$ and $1.5\times10^{-14}/\tau$, respectively. The red dashed line represents the simulation result used to predict the stability of the comparison between ZDT and Sr1, given by $\sqrt{(\sigma_1^2+\sigma_2^2)/2}$. The red symbols represent the Allan deviation calculated by combining all the datasets in (a) using $(\nu_1-\nu_3)/\sqrt{2}\nu_{\mathrm{Sr}}$.
}
\label{fig4}
\end{figure}

Achieving exceptional long-term frequency stability in optical lattice clocks is essential for realizing their full potential in applications requiring unprecedented temporal precision. However, this stability is constrained by fundamental noise sources during measurement cycles. Furthermore, variations in systematic shifts arising from environmental fluctuations or long-term drifts in experimental parameters also degrade the stability. We estimate the long-term stability budget for each clcok~\cite{supp}, quantifying individual contributions of key noise processes including QPN, atomic detection noise, FNC, and the Dick effect (Table~\ref{table1}). Sr1's clock laser references a cavity with lower noise and utilizes extended spectroscopy time to partially reduce the Dick effect, resulting in a stability limit of $9.7\times10^{-17}/\sqrt{\tau}$. The ZDT scheme, by nearly eliminating the Dick effect, which is the dominant noise source in a single clock, achieves a stability limit of $4.1\times10^{-17}/\sqrt{\tau}$, approaching the fundamental limit set by QPN and atomic detection noise, representing an order-of-magnitude improvement over the single-clock stability. We also analyze the systematic shift drifts using recorded data which shows the residual drifts of BBR, Zeeman, and density shifts are at the $10^{-19}$ level or lower for averaging times exceeding 10,000 s.

To comprehensively characterize the stability performance of the ZDT scheme, we performed repeated frequency comparison experiments (Fig.~\ref{fig4}). The dataset consists of four independent ZDT operation cycles, each interrupted by scheduled technical maintenance. The right panel displays the histogram of frequency differences, which exhibits a nearly Gaussian distribution centered at the mean, with no significant outliers or skewness. As the averaging time increases, the overall stability trend aligns well with the simulation prediction, showing a progressive transition from the Sr3 single-clock stability profile toward the limit of the Sr1 clock~\cite{supp}. For averaging times between 4000 s and 20,000 s, the instability degrades proportionally to $1/\tau$. Simulation analysis quantitatively confirms that the onset time of this convergence process are critically governed by the limited feedback bandwidth of atomic servo, characterized by feedback executed every few seconds, and the residual time-dependent frequency drift of the ULE cavities. The combined dataset from four experiments achieves a stability of $2.5\!\times\!10^{-19}$ at 86,400 seconds (1 day), robustly confirming the reproducibility and reliability of the frequency comparison across multiple experimental campaigns. The simulation yields ZDT's one-day stability of $2.0\times10^{-19}$~\cite{supp}.

%%=====================
%% Summary
%%=====================
\section{Summary}\label{sec13}

In summary, this work demonstrates, for the first time, through the comparisons of two independent optical clocks, the superiority of the ZDT architecture in suppressing the Dick effect and accelerating convergence to the stability limit. Specifically, a ZDT stability of ~$2.9 \times 10^{-19}$ is achieved at an averaging time of 20,000~seconds. Extending to 1 day, the combined dataset also predicts stability at the $10^{-19}$ level, confirming the robustness and reproducibility of the comparisons. These results indicate that even with the miniaturized atomic physics system and lower-performance OLO, the ZDT architecture can still achieve superior stability. Further improvements can be realized through refined control of the timing and profile of Ramsey pulses to optimize the uniformity of the composite sensitivity function~\cite{kim2023}. This advancement paves the way for future applications in high-precision portable~\cite{takamoto2020,ohmae2021,bothwell2025} and space-borne optical clocks~\cite{bongs2015,kolkowitz2016,derevianko2022,schkolnik2023}.

\begin{acknowledgments}
The authors would like to thank H.-F. Jiang and R. Shu for valuable discussions.

This work is supported by 
the Scientific Research Innovation Capability Support Project for Young Faculty (Gtant No.~{ZYGXQNJSKYCXNLZCXM-I26}), 
the Anhui Initiative in Quantum Information Technologies (Grant No.~{AHY040200}), 
the National Key Research and Development Program of China (Grant No.~{2020YFA0309804}), 
the Shanghai Municipal Science and Technology Major Project (Grant No.~{2019SHZDZX01}), 
the Innovation Program for Quantum Science and Technology (Grants No.~{2021ZD0302002} and No.~{2021ZD0300106}), 
the Strategic Priority Research Program of Chinese Academy of Sciences (Grant No.~{XDB35000000}), 
and the New Cornerstone Science Foundation.

{\it $^{\#}$ X.-Y. L., P. L. and J. L. contributed equally to this work.}

\end{acknowledgments}

%

%%%%%%%%%%%%%%%%%%%%%%%%%%%%%%%%%%%%%%%%%%%%%%%%
%%%%%%% Supplemental materials %%%%%%%%%%%%%%%%%
%%%%%%%%%%%%%%%%%%%%%%%%%%%%%%%%%%%%%%%%%%%%%%%%
\cleardoublepage
\clearpage
\newpage

\setcounter{equation}{0}
\setcounter{figure}{0}
\setcounter{table}{0}
\setcounter{page}{1}
\setcounter{section}{0}
\makeatletter
\makeatother
\global\def\theequation{S\arabic{equation}}
\global\def\thefigure{S\arabic{figure}}%
\global\def\thetable{S\arabic{table}}
\global\def\thepage{S\arabic{page}}
\global\def\thesection{S\arabic{section}}
\renewcommand{\bibnumfmt}[1]{[S#1]}
\renewcommand{\citenumfont}[1]{S#1}

\onecolumngrid
\begin{center}
	{\large Supplementary material for\\
		\vspace{0.2em}
		``{\bf A zero-dead-time strontium lattice clock with a stability at  $10^{-19}$ level}''
	}\\
	\vspace{0.5em}
	Xiao-Yong~{Liu {\em et. al.}}\\
	\vspace{0.2em}
	(\today)\\
	\vspace{0.5em}
\end{center}

%\twocolumngrid

\section{Experimental Setup}\label{sm-sec1}

The Sr1 system implements three key upgrades relative to prior work~\cite{li2024strontium,Li2023,jia2025uncertainty}. First, a lattice enhancement cavity integrates a piezo ring chip in the output mirror for active length stabilization, where a Pound-Drever-Hall (PDH) lock maintains TEM00 resonance through high-frequency laser current modulation and low-frequency piezo control. Simultaneously, a separate PDH loop phase-locks the laser to a 10-cm ULE cavity (finesse $1.5\times10^4$, drift <1 kHz/day), forming a dual-loop architecture that suppresses relative intensity noise (RIN) while ensuring sub-kHz/day stability; spectrally purified sum-frequency-generation combined with volume Bragg grating filtering~\cite{Jia2025} suppresses background lattice light shifts~\cite{Fasano2021} below $1\times10^{-20}$. Second, magnetically insensitive $\sigma$-transitions $|^{1}\mathrm{S}_0,m_F =\pm \frac{5}{2}\rangle\!\leftrightarrow\!|^{3}\mathrm{P}_0,m_F =\pm\frac{3}{2}\rangle$~\cite{oelker2019} are employed to suppress first-order Zeeman shifts, reducing Zeeman coefficients by 22× compared to stretched-state 
$\pi$-transitions and minimizing magnetic noise coupling to clock frequency. Third, enhanced in-lattice cooling protocols execute simultaneous axial and radial cooling: axial sideband cooling via 689-nm lattice-overlapped laser achieves $n_z=0$ population 0.99 within 20 ms, while radial cooling using three-axis 689-nm beams~\cite{Ushijima2018} operates for 30 ms with optimized parameters to minimize radial temperature while preserving axial ground-state purity; further cooling via energy filtering~\cite{Falke2014}, where the lattice depth undergoes sigmoidal ramping from $200E_\mathrm{r}$ to $41E_\mathrm{r}$ over 35 ms, holds for 35 ms, then ramps to operational depth. 

The Sr3 system implements a miniaturized and integrated architecture featuring a compact atomic source, reduced-scale science chamber, and consolidated supporting optics. Four strontium dispensers are installed within the atomic source vacuum chamber to generate atomic beams. Each dispenser is clamped to vacuum feedthrough electrodes at both ends. When heated beyond 350°C, the dispensers eject atomic flux toward the chamber's geometric center, where a 2D MOT captures low-velocity atoms near the central axis.The timing sequence of actomic state preparation for Sr3 is shown in FIG.~\ref{figS2}. To prepare the $^{87}$Sr cold atomic sample, a two-stage MOT sequence cools the atoms. Atoms are then loaded into a one-dimensional magic-wavelength optical lattice (813.43 nm). The lattice forms via counter-propagating Gaussian beams, confining atoms at standing-wave antinodes where the differential AC Stark shift between clock states is minimized. In the Lamb-Dicke regime, Doppler and photon-recoil shifts from atomic motion are suppressed, enabling precise clock transition measurements. Spin polarization is achieved by optical pumping into the stretched states $|^1S_0, m_F = \pm \frac{9}{2}\rangle$. A 500 mG bias field resolves Zeeman sublevels, and Ramsey spectroscopy interrogates the $|^1S_0, m_F = -\frac{9}{2}\rangle \rightarrow |^3P_0, m_F = -\frac{9}{2}\rangle$ transition. The excitation fraction is determined via electron shelving using a photomultiplier tube (PMT) to detect ground-state and excited-state atom populations. This alternates with an identical sequence on the $|^1S_0, m_F = +\frac{9}{2}\rangle \rightarrow |^3P_0, m_F = +\frac{9}{2}\rangle$ transition. As these transitions exhibit opposite first-order Zeeman shifts, averaging correction signals from both servos suppresses frequency deviations from magnetic field fluctuations in clock stability calculations. This technique effectively mitigates slowly varying magnetic field noise. Magnetic field compensation frequencies are applied to AOM3a(b) for Sr3a(b), and to AOM1a for Sr1.

\begin{sidewaysfigure}
	\centering
	\includegraphics[width=\linewidth]{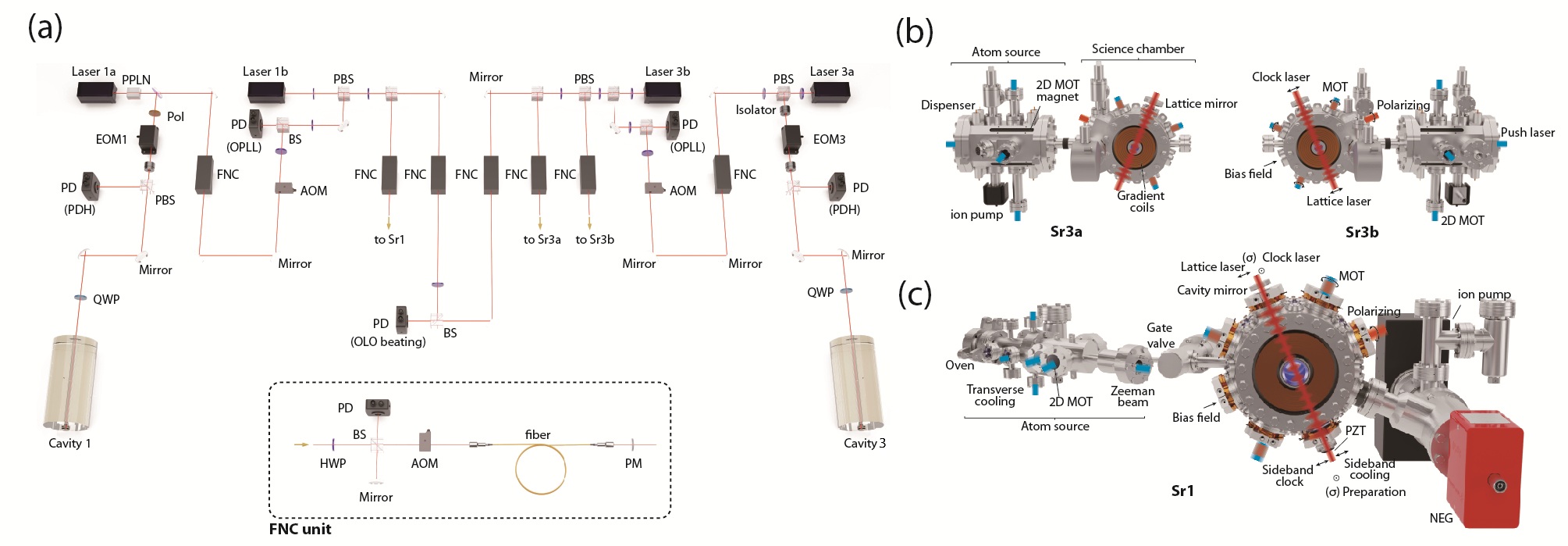}
	\caption{
		Experimental setup. (a) Optical setup for frequency stabilization of the clock lasers referenced to ULE cavities. Abbreviations: Pol, polarizer; HWP, half-wave plate; QWP, quarter-wave plate; PBS, polarization beam splitter; BS, beam splitter; EOM, electro-optic modulator; PD, photodetector; FNC, fiber noise cancellation;  OPLL, optical phase-locked loop;  PM, partial mirror. (b) System architecture and optical layout of Sr3. Structural components are labeled in the left panel, and optical components are labeled in the right panel. Arrows adjacent to each beam indicate its polarization direction. (c) System architecture and optical layout of Sr1. The diagram depicts the vacuum assembly and optical layout, including the configuration of the enhancement cavity for the lattice.
	}
	\label{figS1}
\end{sidewaysfigure}

\begin{figure}[htb]
	\centering
	\includegraphics[width=0.67\linewidth]{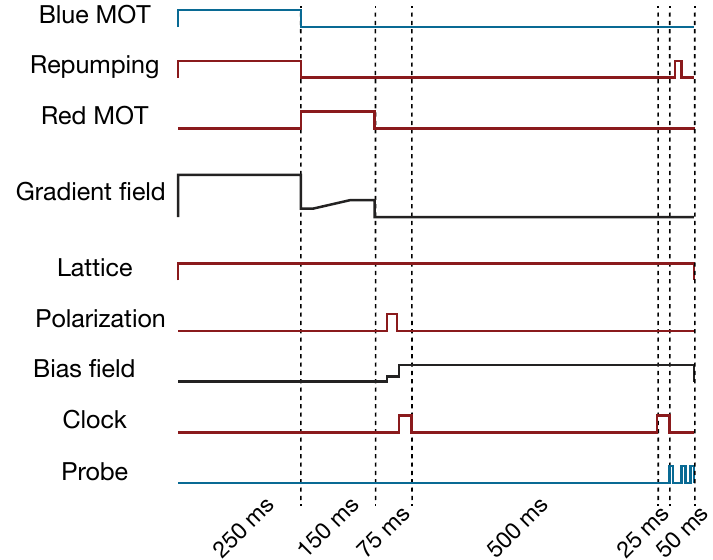}
	\caption{
		The operational timing sequence for Sr3 clock.This diagram depicts the sequence of a single system performing atomic sample preparation, clock transition interrogation, and readout of the excitation.
	}
	\label{figS2}
\end{figure}

The two clocks reside on separate floors, with optical signals for beat-note measurement and phase locking distributed through 70-meter fibers with fiber noise  cancellation (FNC), achieving in-loop fractional stability below $1\times10^{-17}$ at 1 s averaging time. Fig.\ref{figS1} presents the optical configuration for the frequency stabilization of the clock lasers and the detailed apparatus of the two atomic systems.

\section{Performance of OLOs}\label{sm-sec2}

\begin{figure}[htb]
	\centering
	\includegraphics[width=0.67\linewidth]{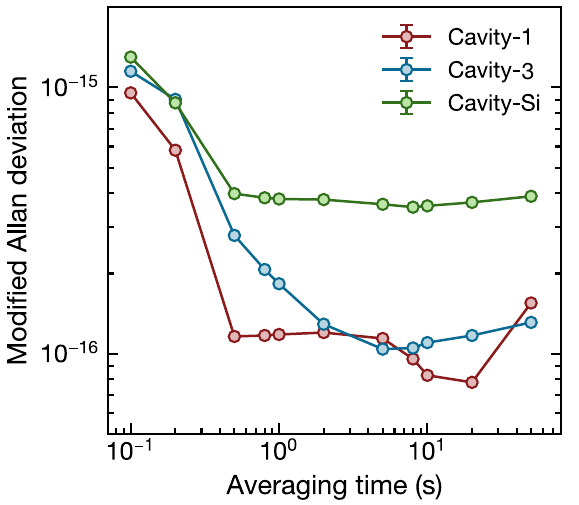}
	\caption{
		Characterization of OLO stability. The fractional frequency instability of the 1397 nm 30-cm-long system Cavity-1 (red), the 698 nm 30-cm-long system Cavity-3 (blue) and the 1397 nm 11.25-cm-long system Cavity-Si (green) is obtained using the TCH method.}
	\label{figS3}
\end{figure}

To assess the noise level of OLOs, we conduct cross spectral density (CSD) measurements.
This method utilizes three photodetectors to capture beat signals between pairs of the three ultrastable lasers referenced on Cavity-1, Cavity-3 and Si1 cavities, where Si1 is an 11.25 cm silicon (Si) cavity operating at 124 K~\cite{zhu2025}. After removing linear drift, the PSD of the three lasers are calculated. The Allan deviation of the ultrastable laser frequencies is derived using the triangle hat (TCH) method applied to the beat signals. During these measurements, the lasers were stabilized solely to the ultrastable cavities without feedback from the atomic clock systems. The TCH analysis demonstrates frequency instabilities below $2\!\times\!10^{-16}$ for the Cavity-1 and Cavity-3 laser systems, at averaging times from 1 to 50 s, as shown in Fig.\ref{figS3}. 	The fractional frequency stability is $1.2\!\times\!10^{-16}$ for Cavity-1 and $1.8\!\times\!10^{-16}$ for Cavity-3 at 1 s, respectively. Specially, the stability for Cavity-1 enters 10$^{-17}$ for averaging times of 10-20 s.

% The CSD measurement results, shown in Fig.\ref{figS2} (a), reveal characteristic peaks near 2.3 Hz with similar amplitudes across all three ultrastable lasers, as well as additional peaks in the 10–50 Hz range with comparable frequencies but varying amplitudes—the largest amplitude observed in cavity-2. These shared spectral features correlate with the common geological and traffic environment of the three ultrastable cavities. The TCH analysis in Fig.\ref{figS1} (b) demonstrates frequency instabilities below $3.1\!\times\!10^{-16}$ and $3.3\!\times\!10^{-16}$ for the cavity-1, cavity-2 laser systems, respectively, at averaging times from 1 to 100 seconds.

\section{Dick effect calculation}\label{sm-sec3}

We can calculate the Dick-effect-induced contribution to clock instability by combining the PSD of OLO frequency noise with the timing sequence. The impact of these noise on clock stability can be minimized by optimizing the dead time and interrogation time settings in the operational timing sequence.
The clock instability arising from the Dick effect is~\cite{dick1989local,oelker2019}
\begin{equation}
	\sigma^2_\mathrm{Dick}=\frac{1}{\tau}\sum_{k=1}^{\infty}\left|\frac{G(k/T_\mathrm{c})}{G(0)}\right|^2 S_y\left(k/T_\mathrm{c}\right).
	\label{eq_dick_noise}
\end{equation}
where $S_y\left(k/T_\mathrm{c}\right)$ is the single-sided power spectral density (PSD) of the fractional frequency fluctuations of the OLO, and $G(f)$ is the atomic sensitivity function at Fourier frequency $f$.

By optimizing the spectroscopy time and dead time of the Sr1 clock, it is possible to achieve a lower Dick effect and thus higher clock stability. The general trend indicates that a larger spectroscopy time and a smaller dead time lead to a reduced Dick effect limit. However, in practical experiments, due to constraints imposed by the laser-atom coherence time and the atom preparation process, we ultimately set the operating parameters to spectroscopy time 1400 ms and dead time 520 ms. Under these conditions, the estimated Dick effect for Sr1 is $9.0\!\times\!10^{-17}/\sqrt{\tau}$. The timing sequence for Sr3a(b) comprises a 500-ms dead time, two 25-ms $pi$/2 pulses, and a 500-ms free evolution period, yielding a Dick instability of $3.8\!\times\!10^{-16}/\sqrt{\tau}$. When operating in ZDT mode, Sr3 achieves a calculated Dick effect instability below $2.8\!\times\!10^{-18}/\sqrt{\tau}$ due to combined sensitivity function of both subsystems.

\section{Simulation of clock stability}\label{sm-sec4}

To interpret the instability results in this work, we simulate the frequency stabilization process of OLOs referenced to atomic transitions. The simulation initiates by generating time-domain frequency noise signals for free-running OLOs through inverse fast Fourier transform (IFFT) of their PSDs. The time-domain signal is computed via $f(t)=\mathcal{F}^{-1}\{\sqrt{S[k]f_s N}e^{i\phi_k}\}$, where $\mathcal{F}^{-1}$ denotes the IFFT, $S[k]$ is the two-sided PSD for each frequency bin k, $f_s$ is the sampling frequency, $N$ is the number of frequency bins, and $\phi_k$ is random phase. By monitoring the frequency values fed to the AOMs during closed-loop process, we obtain the long-term drift of the OLO frequency. Experimentally representative long-term drifts are then incorporated to obtain the signal $f_{1}(t)$, yielding a realistic approximation of laser detuning variations over time. Using this signal, we obtain the detuning of the OLO frequency relative to the atomic transition, thereby deriving the excitation fractions. The correction generated by the employed PID algorithm is compensated into the signal, yielding $f_2(t)$. The Allan deviation of $f_2(t)$ quantifies the stability of the frequency-corrected OLO. 

\begin{figure}[htb]
	\centering
	\includegraphics[width=0.67\linewidth]{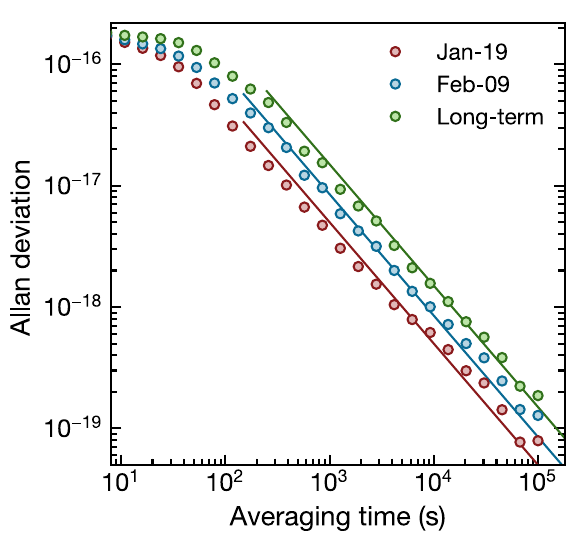}
	\caption{
		Simulated ZDT clock stability. The dots represent the Allan deviation calculated from the simulated time-domain frequency data. The solid lines correspond to fits using a function proportional to $1/\tau$. The resulting expressions for the three fitted lines are $5.0\times10^{-15}/\tau$, $8.5\times10^{-15}/\tau$ and $1.5\times10^{-14}/\tau$, respectively.}
	\label{figS7}
\end{figure}

\begin{figure}[htb]
	\centering
	\includegraphics[width=0.67\linewidth]{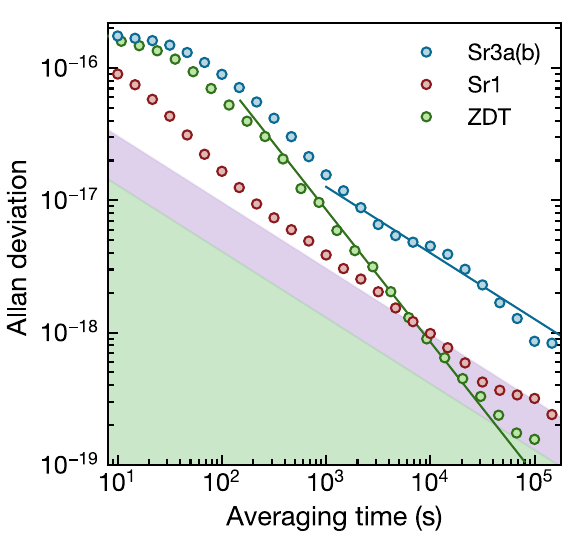}
	\caption{
		Simulation of clock stability.The purple and green shaded regions represent the estimated stability limits of Sr1 and ZDT, respectively. The green line represents the fit for ZDT, yielding a result of $ 8.5\!\times\!10^{-15}/\tau$, while the blue line for Sr3a(b) gives a fit result of $3.9\!\times\!10^{-17}/\sqrt{\tau}$.
	}
	\label{figS4}
\end{figure}

We simulated the ZDT stability under three scenarios using the same algorithm, as shown in Fig.\ref{figS7}. These three scenarios correspond to simulations based on three sets of measured data. During different time periods, the residual frequency drifts and PSDs of the clock laser are different, which is affected by factors such as temperature and vibration. This variation leads to distinct noise levels in the time-domain signals. After feedback correction via ZDT, the Allan deviation exhibits differences in the onset time of its $1/\tau$ decreasing trend, thereby giving rise to the three curves in the figure. For the best result Jan-19 (corresponding to Fig.~2~(d) in the main text), the simulated stability can be fitted as $5.0\times10^{-15}/\tau$. For the evaluation of the long-term stability of ZDT (corresponding to Fig.~4 in the main text), the expectation from the simulation is $1.5\times10^{-14}/\tau$. Fig.\ref{figS4} shows that the simulated stability of each clock system aligns well with our theoretical projections. Specifically, the Allan deviation of Sr1 progressively approaches its fundamental limit of $9.7\!\times\!10^{-17}/\sqrt{\tau}$. Sr3a(b) implementations consistently achieve a baseline stability near $3.9\!\times\!10^{-17}/\sqrt{\tau}$. Meanwhile, the ZDT configuration initially degrades along $ 8.5\!\times\!10^{-15}/\tau$ before asymptotically approaching its ultimate noise floor $4.1\!\times\!10^{-17}/\sqrt{\tau}$, governed by QPN and technical limitations.

% The simulated fractional frequency stability for each clock system exhibit substantial qualitative alignment with our theoretically projected expectations, as shown in Fig.\ref{figS4}. Specifically, the Allan deviation profile of the Sr1 demonstrates progressive convergence toward its fundamental limit of $9.7\!\times\!10^{-17}/\sqrt{\tau}$, while Sr3a(b) implementations consistently manifest stability baselines conforming to $4.0\!\times\!10^{-17}/\sqrt{\tau}$. Concurrently, the ZDT configuration exhibits characteristic stability degradation along the $1.0\!\times\!10^{-14}/\tau$ trajectory during initial operation, subsequently asymptotically approaching its ultimate noise floor governed by QPN and technical limitations.

% We evaluate each clock's instability through simulation analysis. First, we derive the time-domain laser frequency values from the PSD using the following steps:
% \begin{enumerate}
	%     \item Calculate the noise spectrum in the positive frequency range based on the given PSD;
	%     \item Generate random phases to construct the complete noise spectrum;
	%     \item Compute the time-domain noise sequence via inverse Fourier transform;
	%     \item Incorporate the cavity drift calculated from the closed-loop AOM feedback into the time-domain signal, yielding a realistic approximation of laser frequency variations over time.
	% \end{enumerate}

\section{Clock Comparison Results at Different Times}\label{sm-sec5}

\begin{figure*}[htb]
	\centering
	\includegraphics[width=0.67\linewidth]{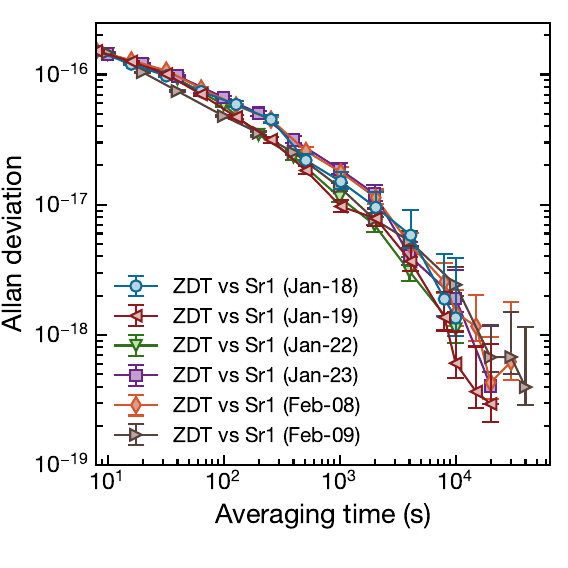}
	\caption{
		Clock comparison results at different times. The experimental sets under ZDT mode were conducted under identical conditions. Among these datasets, the stability of the Jan-19, Jan-23, Feb-08, and Feb-09 datasets reached the $10^{-19}$ level after 10,000 seconds. The other two datasets lacked data beyond this timeframe due to insufficient data collection resulting from system maintenance.}
	\label{figS5}
\end{figure*}

\begin{figure*}[htb]
	\centering
	\includegraphics[width=0.67\linewidth]{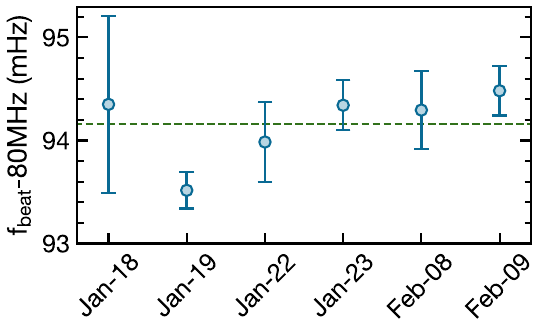}
	\caption{
		Characterization of reproducibility. Each blue dot represents the mean value of each comparison result. Error bars represent 1$\sigma$ uncertainty obtained from the Allan deviation of the two-clock comparisons. The dashed line represents the mean of all data.}
	\label{figS6}
\end{figure*}

To verify the exceptional performance initially observed under the first experimental conditions and rule out potential systematic errors or coincidental factors, we independently repeated the frequency comparison between the Sr1 and Sr3 optical clocks using the identical experimental configuration and measurement protocol, with results shown in Fig.~\ref{figS5}. Across multiple repetitions, the system consistently achieved a frequency stability at the $10^{-19}$ level, with the best-performing result (Jan-19) being used in the main text. This reproducibility robustly demonstrates that the previously reported performance was non-coincidental, providing evidence for the inherent stability of the system and the reliability of our methodology. These results confirm that the Sr1 and Sr3 clocks can reliably attain $10^{-19}$-level frequency stability under the specified operating conditions. Fig. \ref{figS6} presents the frequency difference data obtained from six comparison measurements. The measured frequency differences exhibit variations within ± 1 mHz of the mean, demonstrating minimal systematic drifts and confirming effective control of experimental conditions.

\clearpage
\newpage

\clearpage
\newpage

\clearpage
\newpage

\clearpage
\newpage

\clearpage
\newpage

\clearpage
\newpage

\clearpage
\newpage


\begin{thebibliography}{49}%
	\makeatletter
	\providecommand \@ifxundefined [1]{%
		\@ifx{#1\undefined}
	}%
	\providecommand \@ifnum [1]{%
		\ifnum #1\expandafter \@firstoftwo
		\else \expandafter \@secondoftwo
		\fi
	}%
	\providecommand \@ifx [1]{%
		\ifx #1\expandafter \@firstoftwo
		\else \expandafter \@secondoftwo
		\fi
	}%
	\providecommand \natexlab [1]{#1}%
	\providecommand \enquote  [1]{``#1''}%
	\providecommand \bibnamefont  [1]{#1}%
	\providecommand \bibfnamefont [1]{#1}%
	\providecommand \citenamefont [1]{#1}%
	\providecommand \href@noop [0]{\@secondoftwo}%
	\providecommand \href [0]{\begingroup \@sanitize@url \@href}%
	\providecommand \@href[1]{\@@startlink{#1}\@@href}%
	\providecommand \@@href[1]{\endgroup#1\@@endlink}%
	\providecommand \@sanitize@url [0]{\catcode `\\12\catcode `\$12\catcode
		`\&12\catcode `\#12\catcode `\^12\catcode `\_12\catcode `\%12\relax}%
	\providecommand \@@startlink[1]{}%
	\providecommand \@@endlink[0]{}%
	\providecommand \url  [0]{\begingroup\@sanitize@url \@url }%
	\providecommand \@url [1]{\endgroup\@href {#1}{\urlprefix }}%
	\providecommand \urlprefix  [0]{URL }%
	\providecommand \Eprint [0]{\href }%
	\providecommand \doibase [0]{https://doi.org/}%
	\providecommand \selectlanguage [0]{\@gobble}%
	\providecommand \bibinfo  [0]{\@secondoftwo}%
	\providecommand \bibfield  [0]{\@secondoftwo}%
	\providecommand \translation [1]{[#1]}%
	\providecommand \BibitemOpen [0]{}%
	\providecommand \bibitemStop [0]{}%
	\providecommand \bibitemNoStop [0]{.\EOS\space}%
	\providecommand \EOS [0]{\spacefactor3000\relax}%
	\providecommand \BibitemShut  [1]{\csname bibitem#1\endcsname}%
	\let\auto@bib@innerbib\@empty
	%</preamble>
	\bibitem [{\citenamefont {Ludlow}\ \emph {et~al.}(2015)\citenamefont {Ludlow},
		\citenamefont {Boyd}, \citenamefont {Ye}, \citenamefont {Peik},\ and\
		\citenamefont {Schmidt}}]{ludlow2015optical}%
	\BibitemOpen
	\bibfield  {author} {\bibinfo {author} {\bibfnamefont {A.~D.}\ \bibnamefont
			{Ludlow}}, \bibinfo {author} {\bibfnamefont {M.~M.}\ \bibnamefont {Boyd}},
		\bibinfo {author} {\bibfnamefont {J.}~\bibnamefont {Ye}}, \bibinfo {author}
		{\bibfnamefont {E.}~\bibnamefont {Peik}},\ and\ \bibinfo {author}
		{\bibfnamefont {P.~O.}\ \bibnamefont {Schmidt}},\ }\bibfield  {title}
	{\bibinfo {title} {Optical atomic clocks},\ }\href
	{https://doi.org/10.1103/RevModPhys.87.637} {\bibfield  {journal} {\bibinfo
			{journal} {Reviews of Modern Physics}\ }\textbf {\bibinfo {volume} {87}},\
		\bibinfo {pages} {637} (\bibinfo {year} {2015})}\BibitemShut {NoStop}%
	\bibitem [{\citenamefont {McGrew}\ \emph {et~al.}(2018)\citenamefont {McGrew},
		\citenamefont {Zhang}, \citenamefont {Fasano}, \citenamefont {Sch{\"a}ffer},
		\citenamefont {Beloy}, \citenamefont {Nicolodi}, \citenamefont {Brown},
		\citenamefont {Hinkley}, \citenamefont {Milani}, \citenamefont {Schioppo},
		\citenamefont {Yoon},\ and\ \citenamefont {Ludlow}}]{mcgrew2018}%
	\BibitemOpen
	\bibfield  {author} {\bibinfo {author} {\bibfnamefont {W.~F.}\ \bibnamefont
			{McGrew}}, \bibinfo {author} {\bibfnamefont {X.}~\bibnamefont {Zhang}},
		\bibinfo {author} {\bibfnamefont {R.~J.}\ \bibnamefont {Fasano}}, \bibinfo
		{author} {\bibfnamefont {S.~A.}\ \bibnamefont {Sch{\"a}ffer}}, \bibinfo
		{author} {\bibfnamefont {K.}~\bibnamefont {Beloy}}, \bibinfo {author}
		{\bibfnamefont {D.}~\bibnamefont {Nicolodi}}, \bibinfo {author}
		{\bibfnamefont {R.~C.}\ \bibnamefont {Brown}}, \bibinfo {author}
		{\bibfnamefont {N.}~\bibnamefont {Hinkley}}, \bibinfo {author} {\bibfnamefont
			{G.}~\bibnamefont {Milani}}, \bibinfo {author} {\bibfnamefont
			{M.}~\bibnamefont {Schioppo}}, \bibinfo {author} {\bibfnamefont {T.~H.}\
			\bibnamefont {Yoon}},\ and\ \bibinfo {author} {\bibfnamefont {A.~D.}\
			\bibnamefont {Ludlow}},\ }\bibfield  {title} {\bibinfo {title} {Atomic clock
			performance enabling geodesy below the centimetre level},\ }\href
	{https://doi.org/10.1038/s41586-018-0738-2} {\bibfield  {journal} {\bibinfo
			{journal} {Nature}\ }\textbf {\bibinfo {volume} {564}},\ \bibinfo {pages}
		{87} (\bibinfo {year} {2018})}\BibitemShut {NoStop}%
	\bibitem [{\citenamefont {Aeppli}\ \emph {et~al.}(2024)\citenamefont {Aeppli},
		\citenamefont {Kim}, \citenamefont {Warfield}, \citenamefont {Safronova},\
		and\ \citenamefont {Ye}}]{aeppli2024}%
	\BibitemOpen
	\bibfield  {author} {\bibinfo {author} {\bibfnamefont {A.}~\bibnamefont
			{Aeppli}}, \bibinfo {author} {\bibfnamefont {K.}~\bibnamefont {Kim}},
		\bibinfo {author} {\bibfnamefont {W.}~\bibnamefont {Warfield}}, \bibinfo
		{author} {\bibfnamefont {M.~S.}\ \bibnamefont {Safronova}},\ and\ \bibinfo
		{author} {\bibfnamefont {J.}~\bibnamefont {Ye}},\ }\bibfield  {title}
	{\bibinfo {title} {Clock with $8\times10^{-19}$ {{Systematic Uncertainty}}},\
	}\href {https://doi.org/10.1103/PhysRevLett.133.023401} {\bibfield  {journal}
		{\bibinfo  {journal} {Physical Review Letters}\ }\textbf {\bibinfo {volume}
			{133}},\ \bibinfo {pages} {023401} (\bibinfo {year} {2024})}\BibitemShut
	{NoStop}%
	\bibitem [{\citenamefont {Oelker}\ \emph {et~al.}(2019)\citenamefont {Oelker},
		\citenamefont {Hutson}, \citenamefont {Kennedy}, \citenamefont {Sonderhouse},
		\citenamefont {Bothwell}, \citenamefont {Goban}, \citenamefont {Kedar},
		\citenamefont {Sanner}, \citenamefont {Robinson}, \citenamefont {Marti},
		\citenamefont {Matei}, \citenamefont {Legero}, \citenamefont {Giunta},
		\citenamefont {Holzwarth}, \citenamefont {Riehle}, \citenamefont {Sterr},\
		and\ \citenamefont {Ye}}]{oelker2019}%
	\BibitemOpen
	\bibfield  {author} {\bibinfo {author} {\bibfnamefont {E.}~\bibnamefont
			{Oelker}}, \bibinfo {author} {\bibfnamefont {R.~B.}\ \bibnamefont {Hutson}},
		\bibinfo {author} {\bibfnamefont {C.~J.}\ \bibnamefont {Kennedy}}, \bibinfo
		{author} {\bibfnamefont {L.}~\bibnamefont {Sonderhouse}}, \bibinfo {author}
		{\bibfnamefont {T.}~\bibnamefont {Bothwell}}, \bibinfo {author}
		{\bibfnamefont {A.}~\bibnamefont {Goban}}, \bibinfo {author} {\bibfnamefont
			{D.}~\bibnamefont {Kedar}}, \bibinfo {author} {\bibfnamefont
			{C.}~\bibnamefont {Sanner}}, \bibinfo {author} {\bibfnamefont {J.~M.}\
			\bibnamefont {Robinson}}, \bibinfo {author} {\bibfnamefont {G.~E.}\
			\bibnamefont {Marti}}, \bibinfo {author} {\bibfnamefont {D.~G.}\ \bibnamefont
			{Matei}}, \bibinfo {author} {\bibfnamefont {T.}~\bibnamefont {Legero}},
		\bibinfo {author} {\bibfnamefont {M.}~\bibnamefont {Giunta}}, \bibinfo
		{author} {\bibfnamefont {R.}~\bibnamefont {Holzwarth}}, \bibinfo {author}
		{\bibfnamefont {F.}~\bibnamefont {Riehle}}, \bibinfo {author} {\bibfnamefont
			{U.}~\bibnamefont {Sterr}},\ and\ \bibinfo {author} {\bibfnamefont
			{J.}~\bibnamefont {Ye}},\ }\bibfield  {title} {\bibinfo {title}
		{Demonstration of $4.8\times10^{-17}$ stability at 1 s for two independent
			optical clocks},\ }\href {https://doi.org/10.1038/s41566-019-0493-4}
	{\bibfield  {journal} {\bibinfo  {journal} {Nature Photonics}\ }\textbf
		{\bibinfo {volume} {13}},\ \bibinfo {pages} {714} (\bibinfo {year}
		{2019})}\BibitemShut {NoStop}%
	\bibitem [{\citenamefont {Bothwell}\ \emph {et~al.}(2019)\citenamefont
		{Bothwell}, \citenamefont {Kedar}, \citenamefont {Oelker}, \citenamefont
		{Robinson}, \citenamefont {Bromley}, \citenamefont {Tew}, \citenamefont
		{Ye},\ and\ \citenamefont {Kennedy}}]{bothwell2019}%
	\BibitemOpen
	\bibfield  {author} {\bibinfo {author} {\bibfnamefont {T.}~\bibnamefont
			{Bothwell}}, \bibinfo {author} {\bibfnamefont {D.}~\bibnamefont {Kedar}},
		\bibinfo {author} {\bibfnamefont {E.}~\bibnamefont {Oelker}}, \bibinfo
		{author} {\bibfnamefont {J.~M.}\ \bibnamefont {Robinson}}, \bibinfo {author}
		{\bibfnamefont {S.~L.}\ \bibnamefont {Bromley}}, \bibinfo {author}
		{\bibfnamefont {W.~L.}\ \bibnamefont {Tew}}, \bibinfo {author} {\bibfnamefont
			{J.}~\bibnamefont {Ye}},\ and\ \bibinfo {author} {\bibfnamefont {C.~J.}\
			\bibnamefont {Kennedy}},\ }\bibfield  {title} {\bibinfo {title} {{{JILA SrI}}
			optical lattice clock with uncertainty of $2.0\times10^{-18}$},\ }\href
	{https://doi.org/10.1088/1681-7575/ab4089} {\bibfield  {journal} {\bibinfo
			{journal} {Metrologia}\ }\textbf {\bibinfo {volume} {56}},\ \bibinfo {pages}
		{065004} (\bibinfo {year} {2019})}\BibitemShut {NoStop}%
	\bibitem [{\citenamefont {Marshall}\ \emph {et~al.}(2025)\citenamefont
		{Marshall}, \citenamefont {Castillo}, \citenamefont {{Arthur-Dworschack}},
		\citenamefont {Aeppli}, \citenamefont {Kim}, \citenamefont {Lee},
		\citenamefont {Warfield}, \citenamefont {Hinrichs}, \citenamefont {Nardelli},
		\citenamefont {Fortier}, \citenamefont {Ye}, \citenamefont {Leibrandt},\ and\
		\citenamefont {Hume}}]{marshall2025}%
	\BibitemOpen
	\bibfield  {author} {\bibinfo {author} {\bibfnamefont {M.~C.}\ \bibnamefont
			{Marshall}}, \bibinfo {author} {\bibfnamefont {D.~A.~R.}\ \bibnamefont
			{Castillo}}, \bibinfo {author} {\bibfnamefont {W.~J.}\ \bibnamefont
			{{Arthur-Dworschack}}}, \bibinfo {author} {\bibfnamefont {A.}~\bibnamefont
			{Aeppli}}, \bibinfo {author} {\bibfnamefont {K.}~\bibnamefont {Kim}},
		\bibinfo {author} {\bibfnamefont {D.}~\bibnamefont {Lee}}, \bibinfo {author}
		{\bibfnamefont {W.}~\bibnamefont {Warfield}}, \bibinfo {author}
		{\bibfnamefont {J.}~\bibnamefont {Hinrichs}}, \bibinfo {author}
		{\bibfnamefont {N.~V.}\ \bibnamefont {Nardelli}}, \bibinfo {author}
		{\bibfnamefont {T.~M.}\ \bibnamefont {Fortier}}, \bibinfo {author}
		{\bibfnamefont {J.}~\bibnamefont {Ye}}, \bibinfo {author} {\bibfnamefont
			{D.~R.}\ \bibnamefont {Leibrandt}},\ and\ \bibinfo {author} {\bibfnamefont
			{D.~B.}\ \bibnamefont {Hume}},\ }\bibfield  {title} {\bibinfo {title}
		{High-{{Stability Single-Ion Clock}} with $5.5 \times 10^{- 19}$ {{Systematic
					Uncertainty}}},\ }\href {https://doi.org/10.1103/hb3c-dk28} {\bibfield
		{journal} {\bibinfo  {journal} {Physical Review Letters}\ }\textbf {\bibinfo
			{volume} {135}},\ \bibinfo {pages} {033201} (\bibinfo {year}
		{2025})}\BibitemShut {NoStop}%
	\bibitem [{\citenamefont {Takamoto}\ \emph {et~al.}(2020)\citenamefont
		{Takamoto}, \citenamefont {Ushijima}, \citenamefont {Ohmae}, \citenamefont
		{Yahagi}, \citenamefont {Kokado}, \citenamefont {Shinkai},\ and\
		\citenamefont {Katori}}]{takamoto2020}%
	\BibitemOpen
	\bibfield  {author} {\bibinfo {author} {\bibfnamefont {M.}~\bibnamefont
			{Takamoto}}, \bibinfo {author} {\bibfnamefont {I.}~\bibnamefont {Ushijima}},
		\bibinfo {author} {\bibfnamefont {N.}~\bibnamefont {Ohmae}}, \bibinfo
		{author} {\bibfnamefont {T.}~\bibnamefont {Yahagi}}, \bibinfo {author}
		{\bibfnamefont {K.}~\bibnamefont {Kokado}}, \bibinfo {author} {\bibfnamefont
			{H.}~\bibnamefont {Shinkai}},\ and\ \bibinfo {author} {\bibfnamefont
			{H.}~\bibnamefont {Katori}},\ }\bibfield  {title} {\bibinfo {title} {Test of
			general relativity by a pair of transportable optical lattice clocks},\
	}\href {https://doi.org/10.1038/s41566-020-0619-8} {\bibfield  {journal}
		{\bibinfo  {journal} {Nature Photonics}\ }\textbf {\bibinfo {volume} {14}},\
		\bibinfo {pages} {411} (\bibinfo {year} {2020})}\BibitemShut {NoStop}%
	\bibitem [{\citenamefont {Zhiqiang}\ \emph {et~al.}(2023)\citenamefont
		{Zhiqiang}, \citenamefont {Arnold}, \citenamefont {Kaewuam},\ and\
		\citenamefont {Barrett}}]{Zhang2023}%
	\BibitemOpen
	\bibfield  {author} {\bibinfo {author} {\bibfnamefont {Z.}~\bibnamefont
			{Zhiqiang}}, \bibinfo {author} {\bibfnamefont {K.~J.}\ \bibnamefont
			{Arnold}}, \bibinfo {author} {\bibfnamefont {R.}~\bibnamefont {Kaewuam}},\
		and\ \bibinfo {author} {\bibfnamefont {M.~D.}\ \bibnamefont {Barrett}},\
	}\bibfield  {title} {\bibinfo {title} {${}^{176}\text{Lu}^+$ clock comparison
			at the $10^{-18}$ level via correlation spectroscopy},\ }\href
	{https://doi.org/10.1126/sciadv.adg1971} {\bibfield  {journal} {\bibinfo
			{journal} {Science Advances}\ }\textbf {\bibinfo {volume} {9}},\ \bibinfo
		{pages} {eadg1971} (\bibinfo {year} {2023})}\BibitemShut {NoStop}%
	\bibitem [{\citenamefont {Tofful}\ \emph {et~al.}(2024)\citenamefont {Tofful},
		\citenamefont {Baynham}, \citenamefont {Curtis}, \citenamefont {Parsons},
		\citenamefont {Robertson}, \citenamefont {Schioppo}, \citenamefont {Tunesi},
		\citenamefont {Margolis}, \citenamefont {Hendricks}, \citenamefont {Whale},
		\citenamefont {Thompson},\ and\ \citenamefont {Godun}}]{Tofful2024}%
	\BibitemOpen
	\bibfield  {author} {\bibinfo {author} {\bibfnamefont {A.}~\bibnamefont
			{Tofful}}, \bibinfo {author} {\bibfnamefont {C.~F.~A.}\ \bibnamefont
			{Baynham}}, \bibinfo {author} {\bibfnamefont {E.~A.}\ \bibnamefont {Curtis}},
		\bibinfo {author} {\bibfnamefont {A.~O.}\ \bibnamefont {Parsons}}, \bibinfo
		{author} {\bibfnamefont {B.~I.}\ \bibnamefont {Robertson}}, \bibinfo {author}
		{\bibfnamefont {M.}~\bibnamefont {Schioppo}}, \bibinfo {author}
		{\bibfnamefont {J.}~\bibnamefont {Tunesi}}, \bibinfo {author} {\bibfnamefont
			{H.~S.}\ \bibnamefont {Margolis}}, \bibinfo {author} {\bibfnamefont {R.~J.}\
			\bibnamefont {Hendricks}}, \bibinfo {author} {\bibfnamefont {J.}~\bibnamefont
			{Whale}}, \bibinfo {author} {\bibfnamefont {R.~C.}\ \bibnamefont
			{Thompson}},\ and\ \bibinfo {author} {\bibfnamefont {R.~M.}\ \bibnamefont
			{Godun}},\ }\bibfield  {title} {\bibinfo {title} {${}^{171}\text{Yb}^+$
			optical clock with $2.2\times 10^{-18}$ systematic uncertainty and absolute
			frequency measurements},\ }\href {https://doi.org/10.1088/1681-7575/ad53cd}
	{\bibfield  {journal} {\bibinfo  {journal} {Metrologia}\ }\textbf {\bibinfo
			{volume} {61}},\ \bibinfo {pages} {045001} (\bibinfo {year}
		{2024})}\BibitemShut {NoStop}%
	\bibitem [{\citenamefont {Hausser}\ \emph {et~al.}(2025)\citenamefont
		{Hausser}, \citenamefont {Keller}, \citenamefont {Nordmann}, \citenamefont
		{Bhatt}, \citenamefont {Kiethe}, \citenamefont {Liu}, \citenamefont
		{Richter}, \citenamefont {von Boehn}, \citenamefont {Rahm}, \citenamefont
		{Weyers}, \citenamefont {Benkler}, \citenamefont {Lipphardt}, \citenamefont
		{D\"orscher}, \citenamefont {Stahl}, \citenamefont {Klose}, \citenamefont
		{Lisdat}, \citenamefont {Filzinger}, \citenamefont {Huntemann}, \citenamefont
		{Peik},\ and\ \citenamefont {Mehlst\"aubler}}]{Hausser2025}%
	\BibitemOpen
	\bibfield  {author} {\bibinfo {author} {\bibfnamefont {H.~N.}\ \bibnamefont
			{Hausser}}, \bibinfo {author} {\bibfnamefont {J.}~\bibnamefont {Keller}},
		\bibinfo {author} {\bibfnamefont {T.}~\bibnamefont {Nordmann}}, \bibinfo
		{author} {\bibfnamefont {N.~M.}\ \bibnamefont {Bhatt}}, \bibinfo {author}
		{\bibfnamefont {J.}~\bibnamefont {Kiethe}}, \bibinfo {author} {\bibfnamefont
			{H.}~\bibnamefont {Liu}}, \bibinfo {author} {\bibfnamefont {I.~M.}\
			\bibnamefont {Richter}}, \bibinfo {author} {\bibfnamefont {M.}~\bibnamefont
			{von Boehn}}, \bibinfo {author} {\bibfnamefont {J.}~\bibnamefont {Rahm}},
		\bibinfo {author} {\bibfnamefont {S.}~\bibnamefont {Weyers}}, \bibinfo
		{author} {\bibfnamefont {E.}~\bibnamefont {Benkler}}, \bibinfo {author}
		{\bibfnamefont {B.}~\bibnamefont {Lipphardt}}, \bibinfo {author}
		{\bibfnamefont {S.}~\bibnamefont {D\"orscher}}, \bibinfo {author}
		{\bibfnamefont {K.}~\bibnamefont {Stahl}}, \bibinfo {author} {\bibfnamefont
			{J.}~\bibnamefont {Klose}}, \bibinfo {author} {\bibfnamefont
			{C.}~\bibnamefont {Lisdat}}, \bibinfo {author} {\bibfnamefont
			{M.}~\bibnamefont {Filzinger}}, \bibinfo {author} {\bibfnamefont
			{N.}~\bibnamefont {Huntemann}}, \bibinfo {author} {\bibfnamefont
			{E.}~\bibnamefont {Peik}},\ and\ \bibinfo {author} {\bibfnamefont {T.~E.}\
			\bibnamefont {Mehlst\"aubler}},\ }\bibfield  {title} {\bibinfo {title}
		{${}^{115}\text{In}^{+}\text{-}{}^{172}\text{Yb}^{+}$ coulomb crystal clock
			with $2.5\times10^{-18}$ systematic uncertainty},\ }\href
	{https://doi.org/10.1103/PhysRevLett.134.023201} {\bibfield  {journal}
		{\bibinfo  {journal} {Phys. Rev. Lett.}\ }\textbf {\bibinfo {volume} {134}},\
		\bibinfo {pages} {023201} (\bibinfo {year} {2025})}\BibitemShut {NoStop}%
	\bibitem [{\citenamefont {Zhang}\ \emph {et~al.}(2025)\citenamefont {Zhang},
		\citenamefont {Ma}, \citenamefont {Huang}, \citenamefont {Han}, \citenamefont
		{Hu}, \citenamefont {Wang}, \citenamefont {Zhang}, \citenamefont {Tang},
		\citenamefont {Shi}, \citenamefont {Guan},\ and\ \citenamefont
		{Gao}}]{Zhang2025}%
	\BibitemOpen
	\bibfield  {author} {\bibinfo {author} {\bibfnamefont {B.}~\bibnamefont
			{Zhang}}, \bibinfo {author} {\bibfnamefont {Z.}~\bibnamefont {Ma}}, \bibinfo
		{author} {\bibfnamefont {Y.}~\bibnamefont {Huang}}, \bibinfo {author}
		{\bibfnamefont {H.}~\bibnamefont {Han}}, \bibinfo {author} {\bibfnamefont
			{R.}~\bibnamefont {Hu}}, \bibinfo {author} {\bibfnamefont {Y.}~\bibnamefont
			{Wang}}, \bibinfo {author} {\bibfnamefont {H.}~\bibnamefont {Zhang}},
		\bibinfo {author} {\bibfnamefont {L.}~\bibnamefont {Tang}}, \bibinfo {author}
		{\bibfnamefont {T.}~\bibnamefont {Shi}}, \bibinfo {author} {\bibfnamefont
			{H.}~\bibnamefont {Guan}},\ and\ \bibinfo {author} {\bibfnamefont
			{K.}~\bibnamefont {Gao}},\ }\href {https://arxiv.org/abs/2506.17423}
	{\bibinfo {title} {A liquid-nitrogen-cooled ${}^40\text{Ca}^+$ ion optical
			clock with a systematic uncertainty of $4.6\times 10^{-19}$}} (\bibinfo
	{year} {2025}),\ \Eprint {https://arxiv.org/abs/2506.17423} {arXiv:2506.17423
		[physics.atom-ph]} \BibitemShut {NoStop}%
	\bibitem [{\citenamefont {Schioppo}\ \emph {et~al.}(2017)\citenamefont
		{Schioppo}, \citenamefont {Brown}, \citenamefont {McGrew}, \citenamefont
		{Hinkley}, \citenamefont {Fasano}, \citenamefont {Beloy}, \citenamefont
		{Yoon}, \citenamefont {Milani}, \citenamefont {Nicolodi}, \citenamefont
		{Sherman}, \citenamefont {Phillips}, \citenamefont {Oates},\ and\
		\citenamefont {Ludlow}}]{schioppo2017ultrastable}%
	\BibitemOpen
	\bibfield  {author} {\bibinfo {author} {\bibfnamefont {M.}~\bibnamefont
			{Schioppo}}, \bibinfo {author} {\bibfnamefont {R.~C.}\ \bibnamefont {Brown}},
		\bibinfo {author} {\bibfnamefont {W.~F.}\ \bibnamefont {McGrew}}, \bibinfo
		{author} {\bibfnamefont {N.}~\bibnamefont {Hinkley}}, \bibinfo {author}
		{\bibfnamefont {R.~J.}\ \bibnamefont {Fasano}}, \bibinfo {author}
		{\bibfnamefont {K.}~\bibnamefont {Beloy}}, \bibinfo {author} {\bibfnamefont
			{T.~H.}\ \bibnamefont {Yoon}}, \bibinfo {author} {\bibfnamefont
			{G.}~\bibnamefont {Milani}}, \bibinfo {author} {\bibfnamefont
			{D.}~\bibnamefont {Nicolodi}}, \bibinfo {author} {\bibfnamefont {J.~A.}\
			\bibnamefont {Sherman}}, \bibinfo {author} {\bibfnamefont {N.~B.}\
			\bibnamefont {Phillips}}, \bibinfo {author} {\bibfnamefont {C.~W.}\
			\bibnamefont {Oates}},\ and\ \bibinfo {author} {\bibfnamefont {A.~D.}\
			\bibnamefont {Ludlow}},\ }\bibfield  {title} {\bibinfo {title} {Ultrastable
			optical clock with two cold-atom ensembles},\ }\href
	{https://doi.org/10.1038/nphoton.2016.231} {\bibfield  {journal} {\bibinfo
			{journal} {Nature Photonics}\ }\textbf {\bibinfo {volume} {11}},\ \bibinfo
		{pages} {48} (\bibinfo {year} {2017})}\BibitemShut {NoStop}%
	\bibitem [{\citenamefont {Kim}\ \emph {et~al.}(2023)\citenamefont {Kim},
		\citenamefont {McGrew}, \citenamefont {Nardelli}, \citenamefont {Clements},
		\citenamefont {Hassan}, \citenamefont {Zhang}, \citenamefont {Valencia},
		\citenamefont {Leopardi}, \citenamefont {Hume}, \citenamefont {Fortier},
		\citenamefont {Ludlow},\ and\ \citenamefont {Leibrandt}}]{kim2023}%
	\BibitemOpen
	\bibfield  {author} {\bibinfo {author} {\bibfnamefont {M.~E.}\ \bibnamefont
			{Kim}}, \bibinfo {author} {\bibfnamefont {W.~F.}\ \bibnamefont {McGrew}},
		\bibinfo {author} {\bibfnamefont {N.~V.}\ \bibnamefont {Nardelli}}, \bibinfo
		{author} {\bibfnamefont {E.~R.}\ \bibnamefont {Clements}}, \bibinfo {author}
		{\bibfnamefont {Y.~S.}\ \bibnamefont {Hassan}}, \bibinfo {author}
		{\bibfnamefont {X.}~\bibnamefont {Zhang}}, \bibinfo {author} {\bibfnamefont
			{J.~L.}\ \bibnamefont {Valencia}}, \bibinfo {author} {\bibfnamefont
			{H.}~\bibnamefont {Leopardi}}, \bibinfo {author} {\bibfnamefont {D.~B.}\
			\bibnamefont {Hume}}, \bibinfo {author} {\bibfnamefont {T.~M.}\ \bibnamefont
			{Fortier}}, \bibinfo {author} {\bibfnamefont {A.~D.}\ \bibnamefont
			{Ludlow}},\ and\ \bibinfo {author} {\bibfnamefont {D.~R.}\ \bibnamefont
			{Leibrandt}},\ }\bibfield  {title} {\bibinfo {title} {Improved interspecies
			optical clock comparisons through differential spectroscopy},\ }\href
	{https://doi.org/10.1038/s41567-022-01794-7} {\bibfield  {journal} {\bibinfo
			{journal} {Nature Physics}\ }\textbf {\bibinfo {volume} {19}},\ \bibinfo
		{pages} {25} (\bibinfo {year} {2023})}\BibitemShut {NoStop}%
	\bibitem [{\citenamefont {Nichol}\ \emph {et~al.}(2022)\citenamefont {Nichol},
		\citenamefont {Srinivas}, \citenamefont {Nadlinger}, \citenamefont {Drmota},
		\citenamefont {Main}, \citenamefont {Araneda}, \citenamefont {Ballance},\
		and\ \citenamefont {Lucas}}]{nichol2022}%
	\BibitemOpen
	\bibfield  {author} {\bibinfo {author} {\bibfnamefont {B.~C.}\ \bibnamefont
			{Nichol}}, \bibinfo {author} {\bibfnamefont {R.}~\bibnamefont {Srinivas}},
		\bibinfo {author} {\bibfnamefont {D.~P.}\ \bibnamefont {Nadlinger}}, \bibinfo
		{author} {\bibfnamefont {P.}~\bibnamefont {Drmota}}, \bibinfo {author}
		{\bibfnamefont {D.}~\bibnamefont {Main}}, \bibinfo {author} {\bibfnamefont
			{G.}~\bibnamefont {Araneda}}, \bibinfo {author} {\bibfnamefont {C.~J.}\
			\bibnamefont {Ballance}},\ and\ \bibinfo {author} {\bibfnamefont {D.~M.}\
			\bibnamefont {Lucas}},\ }\bibfield  {title} {\bibinfo {title} {An elementary
			quantum network of entangled optical atomic clocks},\ }\href
	{https://doi.org/10.1038/s41586-022-05088-z} {\bibfield  {journal} {\bibinfo
			{journal} {Nature}\ }\textbf {\bibinfo {volume} {609}},\ \bibinfo {pages}
		{689} (\bibinfo {year} {2022})}\BibitemShut {NoStop}%
	\bibitem [{\citenamefont {Dimarcq}\ \emph {et~al.}(2024)\citenamefont
		{Dimarcq}, \citenamefont {Gertsvolf}, \citenamefont {Mileti}, \citenamefont
		{Bize}, \citenamefont {Oates}, \citenamefont {Peik}, \citenamefont
		{Calonico}, \citenamefont {Ido}, \citenamefont {Tavella}, \citenamefont
		{Meynadier}, \citenamefont {Petit}, \citenamefont {Panfilo}, \citenamefont
		{Bartholomew}, \citenamefont {Defraigne}, \citenamefont {Donley},
		\citenamefont {Hedekvist}, \citenamefont {Sesia}, \citenamefont {Wouters},
		\citenamefont {Dub{\'e}}, \citenamefont {Fang}, \citenamefont {Levi},
		\citenamefont {Lodewyck}, \citenamefont {Margolis}, \citenamefont {Newell},
		\citenamefont {Slyusarev}, \citenamefont {Weyers}, \citenamefont {Uzan},
		\citenamefont {Yasuda}, \citenamefont {Yu}, \citenamefont {Rieck},
		\citenamefont {Schnatz}, \citenamefont {Hanado}, \citenamefont {Fujieda},
		\citenamefont {Pottie}, \citenamefont {Hanssen}, \citenamefont {Malimon},\
		and\ \citenamefont {Ashby}}]{dimarcq2024roadmap}%
	\BibitemOpen
	\bibfield  {author} {\bibinfo {author} {\bibfnamefont {N.}~\bibnamefont
			{Dimarcq}}, \bibinfo {author} {\bibfnamefont {M.}~\bibnamefont {Gertsvolf}},
		\bibinfo {author} {\bibfnamefont {G.}~\bibnamefont {Mileti}}, \bibinfo
		{author} {\bibfnamefont {S.}~\bibnamefont {Bize}}, \bibinfo {author}
		{\bibfnamefont {C.~W.}\ \bibnamefont {Oates}}, \bibinfo {author}
		{\bibfnamefont {E.}~\bibnamefont {Peik}}, \bibinfo {author} {\bibfnamefont
			{D.}~\bibnamefont {Calonico}}, \bibinfo {author} {\bibfnamefont
			{T.}~\bibnamefont {Ido}}, \bibinfo {author} {\bibfnamefont {P.}~\bibnamefont
			{Tavella}}, \bibinfo {author} {\bibfnamefont {F.}~\bibnamefont {Meynadier}},
		\bibinfo {author} {\bibfnamefont {G.}~\bibnamefont {Petit}}, \bibinfo
		{author} {\bibfnamefont {G.}~\bibnamefont {Panfilo}}, \bibinfo {author}
		{\bibfnamefont {J.}~\bibnamefont {Bartholomew}}, \bibinfo {author}
		{\bibfnamefont {P.}~\bibnamefont {Defraigne}}, \bibinfo {author}
		{\bibfnamefont {E.~A.}\ \bibnamefont {Donley}}, \bibinfo {author}
		{\bibfnamefont {P.~O.}\ \bibnamefont {Hedekvist}}, \bibinfo {author}
		{\bibfnamefont {I.}~\bibnamefont {Sesia}}, \bibinfo {author} {\bibfnamefont
			{M.}~\bibnamefont {Wouters}}, \bibinfo {author} {\bibfnamefont
			{P.}~\bibnamefont {Dub{\'e}}}, \bibinfo {author} {\bibfnamefont
			{F.}~\bibnamefont {Fang}}, \bibinfo {author} {\bibfnamefont {F.}~\bibnamefont
			{Levi}}, \bibinfo {author} {\bibfnamefont {J.}~\bibnamefont {Lodewyck}},
		\bibinfo {author} {\bibfnamefont {H.~S.}\ \bibnamefont {Margolis}}, \bibinfo
		{author} {\bibfnamefont {D.}~\bibnamefont {Newell}}, \bibinfo {author}
		{\bibfnamefont {S.}~\bibnamefont {Slyusarev}}, \bibinfo {author}
		{\bibfnamefont {S.}~\bibnamefont {Weyers}}, \bibinfo {author} {\bibfnamefont
			{J.-P.}\ \bibnamefont {Uzan}}, \bibinfo {author} {\bibfnamefont
			{M.}~\bibnamefont {Yasuda}}, \bibinfo {author} {\bibfnamefont {D.-H.}\
			\bibnamefont {Yu}}, \bibinfo {author} {\bibfnamefont {C.}~\bibnamefont
			{Rieck}}, \bibinfo {author} {\bibfnamefont {H.}~\bibnamefont {Schnatz}},
		\bibinfo {author} {\bibfnamefont {Y.}~\bibnamefont {Hanado}}, \bibinfo
		{author} {\bibfnamefont {M.}~\bibnamefont {Fujieda}}, \bibinfo {author}
		{\bibfnamefont {P.-E.}\ \bibnamefont {Pottie}}, \bibinfo {author}
		{\bibfnamefont {J.}~\bibnamefont {Hanssen}}, \bibinfo {author} {\bibfnamefont
			{A.}~\bibnamefont {Malimon}},\ and\ \bibinfo {author} {\bibfnamefont
			{N.}~\bibnamefont {Ashby}},\ }\bibfield  {title} {\bibinfo {title} {Roadmap
			towards the redefinition of the second},\ }\href
	{https://doi.org/10.1088/1681-7575/ad17d2} {\bibfield  {journal} {\bibinfo
			{journal} {Metrologia}\ }\textbf {\bibinfo {volume} {61}},\ \bibinfo {pages}
		{012001} (\bibinfo {year} {2024})}\BibitemShut {NoStop}%
	\bibitem [{\citenamefont {Lodewyck}(2019)}]{lodewyck2019definition}%
	\BibitemOpen
	\bibfield  {author} {\bibinfo {author} {\bibfnamefont {J.}~\bibnamefont
			{Lodewyck}},\ }\bibfield  {title} {\bibinfo {title} {On a definition of the
			{{SI}} second with a set of optical clock transitions},\ }\href
	{https://doi.org/10.1088/1681-7575/ab3a82} {\bibfield  {journal} {\bibinfo
			{journal} {Metrologia}\ }\textbf {\bibinfo {volume} {56}},\ \bibinfo {pages}
		{055009} (\bibinfo {year} {2019})}\BibitemShut {NoStop}%
	\bibitem [{\citenamefont {Riehle}(2015)}]{riehle2015towards}%
	\BibitemOpen
	\bibfield  {author} {\bibinfo {author} {\bibfnamefont {F.}~\bibnamefont
			{Riehle}},\ }\bibfield  {title} {\bibinfo {title} {Towards a redefinition of
			the second based on optical atomic clocks},\ }\href
	{https://doi.org/10.1016/j.crhy.2015.03.012} {\bibfield  {journal} {\bibinfo
			{journal} {Comptes Rendus Physique}\ }\bibinfo {series} {The Measurement of
			Time / {{La}} Mesure Du Temps},\ \textbf {\bibinfo {volume} {16}},\ \bibinfo
		{pages} {506} (\bibinfo {year} {2015})}\BibitemShut {NoStop}%
	\bibitem [{\citenamefont {Huntemann}\ \emph {et~al.}(2014)\citenamefont
		{Huntemann}, \citenamefont {Lipphardt}, \citenamefont {Tamm}, \citenamefont
		{Gerginov}, \citenamefont {Weyers},\ and\ \citenamefont
		{Peik}}]{huntemann2014improved}%
	\BibitemOpen
	\bibfield  {author} {\bibinfo {author} {\bibfnamefont {N.}~\bibnamefont
			{Huntemann}}, \bibinfo {author} {\bibfnamefont {B.}~\bibnamefont
			{Lipphardt}}, \bibinfo {author} {\bibfnamefont {C.}~\bibnamefont {Tamm}},
		\bibinfo {author} {\bibfnamefont {V.}~\bibnamefont {Gerginov}}, \bibinfo
		{author} {\bibfnamefont {S.}~\bibnamefont {Weyers}},\ and\ \bibinfo {author}
		{\bibfnamefont {E.}~\bibnamefont {Peik}},\ }\bibfield  {title} {\bibinfo
		{title} {Improved limit on a temporal variation of {{Mp/Me}} from comparisons
			of {{Yb}}+ and {{Cs}} atomic clocks},\ }\href
	{https://doi.org/10.1103/PhysRevLett.113.210802} {\bibfield  {journal}
		{\bibinfo  {journal} {Physical Review Letters}\ }\textbf {\bibinfo {volume}
			{113}},\ \bibinfo {pages} {210802} (\bibinfo {year} {2014})}\BibitemShut
	{NoStop}%
	\bibitem [{\citenamefont {Bothwell}\ \emph {et~al.}(2022)\citenamefont
		{Bothwell}, \citenamefont {Kennedy}, \citenamefont {Aeppli}, \citenamefont
		{Kedar}, \citenamefont {Robinson}, \citenamefont {Oelker}, \citenamefont
		{Staron},\ and\ \citenamefont {Ye}}]{bothwell2022resolving}%
	\BibitemOpen
	\bibfield  {author} {\bibinfo {author} {\bibfnamefont {T.}~\bibnamefont
			{Bothwell}}, \bibinfo {author} {\bibfnamefont {C.~J.}\ \bibnamefont
			{Kennedy}}, \bibinfo {author} {\bibfnamefont {A.}~\bibnamefont {Aeppli}},
		\bibinfo {author} {\bibfnamefont {D.}~\bibnamefont {Kedar}}, \bibinfo
		{author} {\bibfnamefont {J.~M.}\ \bibnamefont {Robinson}}, \bibinfo {author}
		{\bibfnamefont {E.}~\bibnamefont {Oelker}}, \bibinfo {author} {\bibfnamefont
			{A.}~\bibnamefont {Staron}},\ and\ \bibinfo {author} {\bibfnamefont
			{J.}~\bibnamefont {Ye}},\ }\bibfield  {title} {\bibinfo {title} {Resolving
			the gravitational redshift across a millimetre-scale atomic sample},\ }\href
	{https://doi.org/10.1038/s41586-021-04349-7} {\bibfield  {journal} {\bibinfo
			{journal} {Nature}\ }\textbf {\bibinfo {volume} {602}},\ \bibinfo {pages}
		{420} (\bibinfo {year} {2022})}\BibitemShut {NoStop}%
	\bibitem [{\citenamefont {Naro{\.z}nik}\ \emph {et~al.}(2023)\citenamefont
		{Naro{\.z}nik}, \citenamefont {Bober},\ and\ \citenamefont
		{Zawada}}]{naroznik2023}%
	\BibitemOpen
	\bibfield  {author} {\bibinfo {author} {\bibfnamefont {M.}~\bibnamefont
			{Naro{\.z}nik}}, \bibinfo {author} {\bibfnamefont {M.}~\bibnamefont
			{Bober}},\ and\ \bibinfo {author} {\bibfnamefont {M.}~\bibnamefont
			{Zawada}},\ }\bibfield  {title} {\bibinfo {title} {Ultra-stable optical clock
			cavities as resonant mass gravitational wave detectors in search for new
			physics},\ }\href {https://doi.org/10.1016/j.physletb.2023.138260} {\bibfield
		{journal} {\bibinfo  {journal} {Physics Letters B}\ }\textbf {\bibinfo
			{volume} {846}},\ \bibinfo {pages} {138260} (\bibinfo {year}
		{2023})}\BibitemShut {NoStop}%
	\bibitem [{\citenamefont {{KAGRA Collaboration, LIGO Scientific Collaboration
				and Virgo Collaboration}}(2020)}]{abbott2020}%
	\BibitemOpen
	\bibfield  {author} {\bibinfo {author} {\bibnamefont {{KAGRA Collaboration,
					LIGO Scientific Collaboration and Virgo Collaboration}}},\ }\bibfield
	{title} {\bibinfo {title} {Prospects for observing and localizing
			gravitational-wave transients with {{Advanced LIGO}}, {{Advanced Virgo}} and
			{{KAGRA}}},\ }\href {https://doi.org/10.1007/s41114-020-00026-9} {\bibfield
		{journal} {\bibinfo  {journal} {Living Reviews in Relativity}\ }\textbf
		{\bibinfo {volume} {23}},\ \bibinfo {pages} {3} (\bibinfo {year}
		{2020})}\BibitemShut {NoStop}%
	\bibitem [{\citenamefont {Su}\ \emph {et~al.}(2018)\citenamefont {Su},
		\citenamefont {Wang}, \citenamefont {Wang},\ and\ \citenamefont
		{Jetzer}}]{su2018}%
	\BibitemOpen
	\bibfield  {author} {\bibinfo {author} {\bibfnamefont {J.}~\bibnamefont
			{Su}}, \bibinfo {author} {\bibfnamefont {Q.}~\bibnamefont {Wang}}, \bibinfo
		{author} {\bibfnamefont {Q.}~\bibnamefont {Wang}},\ and\ \bibinfo {author}
		{\bibfnamefont {P.}~\bibnamefont {Jetzer}},\ }\bibfield  {title} {\bibinfo
		{title} {Low-frequency gravitational wave detection via double optical clocks
			in space},\ }\href {https://doi.org/10.1088/1361-6382/aab2eb} {\bibfield
		{journal} {\bibinfo  {journal} {Classical and Quantum Gravity}\ }\textbf
		{\bibinfo {volume} {35}},\ \bibinfo {pages} {085010} (\bibinfo {year}
		{2018})}\BibitemShut {NoStop}%
	\bibitem [{\citenamefont {Kolkowitz}\ \emph {et~al.}(2016)\citenamefont
		{Kolkowitz}, \citenamefont {Pikovski}, \citenamefont {Langellier},
		\citenamefont {Lukin}, \citenamefont {Walsworth},\ and\ \citenamefont
		{Ye}}]{kolkowitz2016}%
	\BibitemOpen
	\bibfield  {author} {\bibinfo {author} {\bibfnamefont {S.}~\bibnamefont
			{Kolkowitz}}, \bibinfo {author} {\bibfnamefont {I.}~\bibnamefont {Pikovski}},
		\bibinfo {author} {\bibfnamefont {N.}~\bibnamefont {Langellier}}, \bibinfo
		{author} {\bibfnamefont {M.~D.}\ \bibnamefont {Lukin}}, \bibinfo {author}
		{\bibfnamefont {R.~L.}\ \bibnamefont {Walsworth}},\ and\ \bibinfo {author}
		{\bibfnamefont {J.}~\bibnamefont {Ye}},\ }\bibfield  {title} {\bibinfo
		{title} {Gravitational wave detection with optical lattice atomic clocks},\
	}\href {https://doi.org/10.1103/PhysRevD.94.124043} {\bibfield  {journal}
		{\bibinfo  {journal} {Physical Review D}\ }\textbf {\bibinfo {volume} {94}},\
		\bibinfo {pages} {124043} (\bibinfo {year} {2016})}\BibitemShut {NoStop}%
	\bibitem [{\citenamefont {Filzinger}\ \emph {et~al.}(2025)\citenamefont
		{Filzinger}, \citenamefont {Caddell}, \citenamefont {Jani}, \citenamefont
		{Steinel}, \citenamefont {Giani}, \citenamefont {Huntemann},\ and\
		\citenamefont {Roberts}}]{filzinger2025}%
	\BibitemOpen
	\bibfield  {author} {\bibinfo {author} {\bibfnamefont {M.}~\bibnamefont
			{Filzinger}}, \bibinfo {author} {\bibfnamefont {A.~R.}\ \bibnamefont
			{Caddell}}, \bibinfo {author} {\bibfnamefont {D.}~\bibnamefont {Jani}},
		\bibinfo {author} {\bibfnamefont {M.}~\bibnamefont {Steinel}}, \bibinfo
		{author} {\bibfnamefont {L.}~\bibnamefont {Giani}}, \bibinfo {author}
		{\bibfnamefont {N.}~\bibnamefont {Huntemann}},\ and\ \bibinfo {author}
		{\bibfnamefont {B.~M.}\ \bibnamefont {Roberts}},\ }\bibfield  {title}
	{\bibinfo {title} {Ultralight {{Dark Matter Search}} with {{Space-Time
					Separated Atomic Clocks}} and {{Cavities}}},\ }\href
	{https://doi.org/10.1103/PhysRevLett.134.031001} {\bibfield  {journal}
		{\bibinfo  {journal} {Physical Review Letters}\ }\textbf {\bibinfo {volume}
			{134}},\ \bibinfo {pages} {031001} (\bibinfo {year} {2025})}\BibitemShut
	{NoStop}%
	\bibitem [{\citenamefont {Arvanitaki}\ \emph {et~al.}(2015)\citenamefont
		{Arvanitaki}, \citenamefont {Huang},\ and\ \citenamefont
		{Van~Tilburg}}]{arvanitaki2015}%
	\BibitemOpen
	\bibfield  {author} {\bibinfo {author} {\bibfnamefont {A.}~\bibnamefont
			{Arvanitaki}}, \bibinfo {author} {\bibfnamefont {J.}~\bibnamefont {Huang}},\
		and\ \bibinfo {author} {\bibfnamefont {K.}~\bibnamefont {Van~Tilburg}},\
	}\bibfield  {title} {\bibinfo {title} {Searching for dilaton dark matter with
			atomic clocks},\ }\href {https://doi.org/10.1103/PhysRevD.91.015015}
	{\bibfield  {journal} {\bibinfo  {journal} {Physical Review D}\ }\textbf
		{\bibinfo {volume} {91}},\ \bibinfo {pages} {015015} (\bibinfo {year}
		{2015})}\BibitemShut {NoStop}%
	\bibitem [{\citenamefont {Itano}\ \emph {et~al.}(1993)\citenamefont {Itano},
		\citenamefont {Bergquist}, \citenamefont {Bollinger}, \citenamefont
		{Gilligan}, \citenamefont {Heinzen}, \citenamefont {Moore}, \citenamefont
		{Raizen},\ and\ \citenamefont {Wineland}}]{itano1993}%
	\BibitemOpen
	\bibfield  {author} {\bibinfo {author} {\bibfnamefont {W.~M.}\ \bibnamefont
			{Itano}}, \bibinfo {author} {\bibfnamefont {J.~C.}\ \bibnamefont
			{Bergquist}}, \bibinfo {author} {\bibfnamefont {J.~J.}\ \bibnamefont
			{Bollinger}}, \bibinfo {author} {\bibfnamefont {J.~M.}\ \bibnamefont
			{Gilligan}}, \bibinfo {author} {\bibfnamefont {D.~J.}\ \bibnamefont
			{Heinzen}}, \bibinfo {author} {\bibfnamefont {F.~L.}\ \bibnamefont {Moore}},
		\bibinfo {author} {\bibfnamefont {M.~G.}\ \bibnamefont {Raizen}},\ and\
		\bibinfo {author} {\bibfnamefont {D.~J.}\ \bibnamefont {Wineland}},\
	}\bibfield  {title} {\bibinfo {title} {Quantum projection noise:
			{{Population}} fluctuations in two-level systems},\ }\href
	{https://doi.org/10.1103/PhysRevA.47.3554} {\bibfield  {journal} {\bibinfo
			{journal} {Physical Review A}\ }\textbf {\bibinfo {volume} {47}},\ \bibinfo
		{pages} {3554} (\bibinfo {year} {1993})}\BibitemShut {NoStop}%
	\bibitem [{\citenamefont {Robinson}\ \emph {et~al.}(2024)\citenamefont
		{Robinson}, \citenamefont {Miklos}, \citenamefont {Tso}, \citenamefont
		{Kennedy}, \citenamefont {Bothwell}, \citenamefont {Kedar}, \citenamefont
		{Thompson},\ and\ \citenamefont {Ye}}]{robi2024}%
	\BibitemOpen
	\bibfield  {author} {\bibinfo {author} {\bibfnamefont {J.~M.}\ \bibnamefont
			{Robinson}}, \bibinfo {author} {\bibfnamefont {M.}~\bibnamefont {Miklos}},
		\bibinfo {author} {\bibfnamefont {Y.~M.}\ \bibnamefont {Tso}}, \bibinfo
		{author} {\bibfnamefont {C.~J.}\ \bibnamefont {Kennedy}}, \bibinfo {author}
		{\bibfnamefont {T.}~\bibnamefont {Bothwell}}, \bibinfo {author}
		{\bibfnamefont {D.}~\bibnamefont {Kedar}}, \bibinfo {author} {\bibfnamefont
			{J.~K.}\ \bibnamefont {Thompson}},\ and\ \bibinfo {author} {\bibfnamefont
			{J.}~\bibnamefont {Ye}},\ }\bibfield  {title} {\bibinfo {title} {Direct
			comparison of two spin-squeezed optical clock ensembles at the $10^{-17}$
			level},\ }\href {https://doi.org/10.1038/s41567-023-02310-1} {\bibfield
		{journal} {\bibinfo  {journal} {Nature Physics}\ }\textbf {\bibinfo {volume}
			{20}},\ \bibinfo {pages} {208} (\bibinfo {year} {2024})}\BibitemShut
	{NoStop}%
	\bibitem [{\citenamefont {Cao}\ \emph {et~al.}(2024)\citenamefont {Cao},
		\citenamefont {Eckner}, \citenamefont {Lukin~Yelin}, \citenamefont {Young},
		\citenamefont {Jandura}, \citenamefont {Yan}, \citenamefont {Kim},
		\citenamefont {Pupillo}, \citenamefont {Ye}, \citenamefont {Darkwah~Oppong},\
		and\ \citenamefont {Kaufman}}]{cao2024}%
	\BibitemOpen
	\bibfield  {author} {\bibinfo {author} {\bibfnamefont {A.}~\bibnamefont
			{Cao}}, \bibinfo {author} {\bibfnamefont {W.~J.}\ \bibnamefont {Eckner}},
		\bibinfo {author} {\bibfnamefont {T.}~\bibnamefont {Lukin~Yelin}}, \bibinfo
		{author} {\bibfnamefont {A.~W.}\ \bibnamefont {Young}}, \bibinfo {author}
		{\bibfnamefont {S.}~\bibnamefont {Jandura}}, \bibinfo {author} {\bibfnamefont
			{L.}~\bibnamefont {Yan}}, \bibinfo {author} {\bibfnamefont {K.}~\bibnamefont
			{Kim}}, \bibinfo {author} {\bibfnamefont {G.}~\bibnamefont {Pupillo}},
		\bibinfo {author} {\bibfnamefont {J.}~\bibnamefont {Ye}}, \bibinfo {author}
		{\bibfnamefont {N.}~\bibnamefont {Darkwah~Oppong}},\ and\ \bibinfo {author}
		{\bibfnamefont {A.~M.}\ \bibnamefont {Kaufman}},\ }\bibfield  {title}
	{\bibinfo {title} {Multi-qubit gates and {{Schr{\"o}dinger}} cat states in an
			optical clock},\ }\href {https://doi.org/10.1038/s41586-024-07913-z}
	{\bibfield  {journal} {\bibinfo  {journal} {Nature}\ }\textbf {\bibinfo
			{volume} {634}},\ \bibinfo {pages} {315} (\bibinfo {year}
		{2024})}\BibitemShut {NoStop}%
	\bibitem [{\citenamefont {Dick}(2337)}]{dick1989local}%
	\BibitemOpen
	\bibfield  {author} {\bibinfo {author} {\bibfnamefont {G.~J.}\ \bibnamefont
			{Dick}},\ }\bibfield  {title} {\bibinfo {title} {Local oscillator induced
			instabilities in trapped ion frequency standards},\ }in\ \href
	{http://www.ion.org/publications/abstract.cfm?jp=p&articleID=15462} {\emph
		{\bibinfo {booktitle} {Proceedings of the 19th {{Annual Precise Time}} and
				{{Time Interval Systems}} and {{Applications Meeting}}}}}\ (\bibinfo {year}
	{1989/12/337})\ pp.\ \bibinfo {pages} {133--147}\BibitemShut {NoStop}%
	\bibitem [{\citenamefont {Parke}\ and\ \citenamefont
		{Schioppo}(2025)}]{parke2025}%
	\BibitemOpen
	\bibfield  {author} {\bibinfo {author} {\bibfnamefont {A.~L.}\ \bibnamefont
			{Parke}}\ and\ \bibinfo {author} {\bibfnamefont {M.}~\bibnamefont
			{Schioppo}},\ }\bibfield  {title} {\bibinfo {title} {Three hundred
			microsecond optical cavity storage time and $10^{-7}$ active {{RAM}}
			cancellation for $10^{-19}$ laser frequency stabilization},\ }\href
	{https://doi.org/10.1364/OL.560815} {\bibfield  {journal} {\bibinfo
			{journal} {Optics Letters}\ }\textbf {\bibinfo {volume} {50}},\ \bibinfo
		{pages} {3405} (\bibinfo {year} {2025})}\BibitemShut {NoStop}%
	\bibitem [{\citenamefont {Kedar}\ \emph {et~al.}(2023)\citenamefont {Kedar},
		\citenamefont {Yu}, \citenamefont {Oelker}, \citenamefont {Staron},
		\citenamefont {Milner}, \citenamefont {Robinson}, \citenamefont {Legero},
		\citenamefont {Riehle}, \citenamefont {Sterr},\ and\ \citenamefont
		{Ye}}]{kedar2023}%
	\BibitemOpen
	\bibfield  {author} {\bibinfo {author} {\bibfnamefont {D.}~\bibnamefont
			{Kedar}}, \bibinfo {author} {\bibfnamefont {J.}~\bibnamefont {Yu}}, \bibinfo
		{author} {\bibfnamefont {E.}~\bibnamefont {Oelker}}, \bibinfo {author}
		{\bibfnamefont {A.}~\bibnamefont {Staron}}, \bibinfo {author} {\bibfnamefont
			{W.~R.}\ \bibnamefont {Milner}}, \bibinfo {author} {\bibfnamefont {J.~M.}\
			\bibnamefont {Robinson}}, \bibinfo {author} {\bibfnamefont {T.}~\bibnamefont
			{Legero}}, \bibinfo {author} {\bibfnamefont {F.}~\bibnamefont {Riehle}},
		\bibinfo {author} {\bibfnamefont {U.}~\bibnamefont {Sterr}},\ and\ \bibinfo
		{author} {\bibfnamefont {J.}~\bibnamefont {Ye}},\ }\bibfield  {title}
	{\bibinfo {title} {Frequency stability of cryogenic silicon cavities with
			semiconductor crystalline coatings},\ }\href
	{https://doi.org/10.1364/OPTICA.479462} {\bibfield  {journal} {\bibinfo
			{journal} {Optica}\ }\textbf {\bibinfo {volume} {10}},\ \bibinfo {pages}
		{464} (\bibinfo {year} {2023})}\BibitemShut {NoStop}%
	\bibitem [{\citenamefont {Yu}\ \emph {et~al.}(2023)\citenamefont {Yu},
		\citenamefont {H{\"a}fner}, \citenamefont {Legero}, \citenamefont {Herbers},
		\citenamefont {Nicolodi}, \citenamefont {Ma}, \citenamefont {Riehle},
		\citenamefont {Sterr}, \citenamefont {Kedar}, \citenamefont {Robinson},
		\citenamefont {Oelker},\ and\ \citenamefont {Ye}}]{yu2023}%
	\BibitemOpen
	\bibfield  {author} {\bibinfo {author} {\bibfnamefont {J.}~\bibnamefont
			{Yu}}, \bibinfo {author} {\bibfnamefont {S.}~\bibnamefont {H{\"a}fner}},
		\bibinfo {author} {\bibfnamefont {T.}~\bibnamefont {Legero}}, \bibinfo
		{author} {\bibfnamefont {S.}~\bibnamefont {Herbers}}, \bibinfo {author}
		{\bibfnamefont {D.}~\bibnamefont {Nicolodi}}, \bibinfo {author}
		{\bibfnamefont {C.~Y.}\ \bibnamefont {Ma}}, \bibinfo {author} {\bibfnamefont
			{F.}~\bibnamefont {Riehle}}, \bibinfo {author} {\bibfnamefont
			{U.}~\bibnamefont {Sterr}}, \bibinfo {author} {\bibfnamefont
			{D.}~\bibnamefont {Kedar}}, \bibinfo {author} {\bibfnamefont {J.~M.}\
			\bibnamefont {Robinson}}, \bibinfo {author} {\bibfnamefont {E.}~\bibnamefont
			{Oelker}},\ and\ \bibinfo {author} {\bibfnamefont {J.}~\bibnamefont {Ye}},\
	}\bibfield  {title} {\bibinfo {title} {Excess {{Noise}} and {{Photoinduced
					Effects}} in {{Highly Reflective Crystalline Mirror Coatings}}},\ }\href
	{https://doi.org/10.1103/PhysRevX.13.041002} {\bibfield  {journal} {\bibinfo
			{journal} {Physical Review X}\ }\textbf {\bibinfo {volume} {13}},\ \bibinfo
		{pages} {041002} (\bibinfo {year} {2023})}\BibitemShut {NoStop}%
	\bibitem [{\citenamefont {Robinson}\ \emph {et~al.}(2019)\citenamefont
		{Robinson}, \citenamefont {Oelker}, \citenamefont {Milner}, \citenamefont
		{Zhang}, \citenamefont {Legero}, \citenamefont {Matei}, \citenamefont
		{Riehle}, \citenamefont {Sterr},\ and\ \citenamefont {Ye}}]{robinson2019}%
	\BibitemOpen
	\bibfield  {author} {\bibinfo {author} {\bibfnamefont {J.~M.}\ \bibnamefont
			{Robinson}}, \bibinfo {author} {\bibfnamefont {E.}~\bibnamefont {Oelker}},
		\bibinfo {author} {\bibfnamefont {W.~R.}\ \bibnamefont {Milner}}, \bibinfo
		{author} {\bibfnamefont {W.}~\bibnamefont {Zhang}}, \bibinfo {author}
		{\bibfnamefont {T.}~\bibnamefont {Legero}}, \bibinfo {author} {\bibfnamefont
			{D.~G.}\ \bibnamefont {Matei}}, \bibinfo {author} {\bibfnamefont
			{F.}~\bibnamefont {Riehle}}, \bibinfo {author} {\bibfnamefont
			{U.}~\bibnamefont {Sterr}},\ and\ \bibinfo {author} {\bibfnamefont
			{J.}~\bibnamefont {Ye}},\ }\bibfield  {title} {\bibinfo {title} {Crystalline
			optical cavity at 4 {{K}} with thermal-noise-limited instability and ultralow
			drift},\ }\href {https://doi.org/10.1364/OPTICA.6.000240} {\bibfield
		{journal} {\bibinfo  {journal} {Optica}\ }\textbf {\bibinfo {volume} {6}},\
		\bibinfo {pages} {240} (\bibinfo {year} {2019})}\BibitemShut {NoStop}%
	\bibitem [{\citenamefont {Katori}(2021)}]{katori2021}%
	\BibitemOpen
	\bibfield  {author} {\bibinfo {author} {\bibfnamefont {H.}~\bibnamefont
			{Katori}},\ }\bibfield  {title} {\bibinfo {title} {Longitudinal {{Ramsey}}
			spectroscopy of atoms for continuous operation of optical clocks},\ }\href
	{https://doi.org/10.35848/1882-0786/ac0e16} {\bibfield  {journal} {\bibinfo
			{journal} {Applied Physics Express}\ }\textbf {\bibinfo {volume} {14}},\
		\bibinfo {pages} {072006} (\bibinfo {year} {2021})}\BibitemShut {NoStop}%
	\bibitem [{\citenamefont {Biedermann}\ \emph {et~al.}(2013)\citenamefont
		{Biedermann}, \citenamefont {Takase}, \citenamefont {Wu}, \citenamefont
		{Deslauriers}, \citenamefont {Roy},\ and\ \citenamefont
		{Kasevich}}]{biedermann2013}%
	\BibitemOpen
	\bibfield  {author} {\bibinfo {author} {\bibfnamefont {G.~W.}\ \bibnamefont
			{Biedermann}}, \bibinfo {author} {\bibfnamefont {K.}~\bibnamefont {Takase}},
		\bibinfo {author} {\bibfnamefont {X.}~\bibnamefont {Wu}}, \bibinfo {author}
		{\bibfnamefont {L.}~\bibnamefont {Deslauriers}}, \bibinfo {author}
		{\bibfnamefont {S.}~\bibnamefont {Roy}},\ and\ \bibinfo {author}
		{\bibfnamefont {M.~A.}\ \bibnamefont {Kasevich}},\ }\bibfield  {title}
	{\bibinfo {title} {Zero-{{Dead-Time Operation}} of {{Interleaved Atomic
					Clocks}}},\ }\href {https://doi.org/10.1103/PhysRevLett.111.170802}
	{\bibfield  {journal} {\bibinfo  {journal} {Physical Review Letters}\
		}\textbf {\bibinfo {volume} {111}},\ \bibinfo {pages} {170802} (\bibinfo
		{year} {2013})}\BibitemShut {NoStop}%
	\bibitem [{\citenamefont {Lin}\ \emph {et~al.}(2017)\citenamefont {Lin},
		\citenamefont {Lin}, \citenamefont {Deng}, \citenamefont {Zhang},\ and\
		\citenamefont {Wang}}]{lin2017}%
	\BibitemOpen
	\bibfield  {author} {\bibinfo {author} {\bibfnamefont {H.}~\bibnamefont
			{Lin}}, \bibinfo {author} {\bibfnamefont {J.}~\bibnamefont {Lin}}, \bibinfo
		{author} {\bibfnamefont {J.}~\bibnamefont {Deng}}, \bibinfo {author}
		{\bibfnamefont {S.}~\bibnamefont {Zhang}},\ and\ \bibinfo {author}
		{\bibfnamefont {Y.}~\bibnamefont {Wang}},\ }\bibfield  {title} {\bibinfo
		{title} {Pulsed optically pumped atomic clock with zero-dead-time},\ }\href
	{https://doi.org/10.1063/1.5008627} {\bibfield  {journal} {\bibinfo
			{journal} {Review of Scientific Instruments}\ }\textbf {\bibinfo {volume}
			{88}},\ \bibinfo {pages} {123103} (\bibinfo {year} {2017})}\BibitemShut
	{NoStop}%
	\bibitem [{\citenamefont {Cheng}\ \emph {et~al.}(2018)\citenamefont {Cheng},
		\citenamefont {Sun}, \citenamefont {Zhang},\ and\ \citenamefont
		{Wang}}]{cheng2018}%
	\BibitemOpen
	\bibfield  {author} {\bibinfo {author} {\bibfnamefont {P.}~\bibnamefont
			{Cheng}}, \bibinfo {author} {\bibfnamefont {X.}~\bibnamefont {Sun}}, \bibinfo
		{author} {\bibfnamefont {J.}~\bibnamefont {Zhang}},\ and\ \bibinfo {author}
		{\bibfnamefont {L.}~\bibnamefont {Wang}},\ }\bibfield  {title} {\bibinfo
		{title} {Suppression of {{Dick Effect}} in {{Ramsey-CPT Atomic Clock}} by
			{{Interleaving Lock}}},\ }\href {https://doi.org/10.1109/TUFFC.2018.2864622}
	{\bibfield  {journal} {\bibinfo  {journal} {IEEE Transactions on Ultrasonics,
				Ferroelectrics, and Frequency Control}\ }\textbf {\bibinfo {volume} {65}},\
		\bibinfo {pages} {2195} (\bibinfo {year} {2018})}\BibitemShut {NoStop}%
	\bibitem [{\citenamefont {Takamoto}\ \emph {et~al.}(2011)\citenamefont
		{Takamoto}, \citenamefont {Takano},\ and\ \citenamefont
		{Katori}}]{takamoto2011}%
	\BibitemOpen
	\bibfield  {author} {\bibinfo {author} {\bibfnamefont {M.}~\bibnamefont
			{Takamoto}}, \bibinfo {author} {\bibfnamefont {T.}~\bibnamefont {Takano}},\
		and\ \bibinfo {author} {\bibfnamefont {H.}~\bibnamefont {Katori}},\
	}\bibfield  {title} {\bibinfo {title} {Frequency comparison of optical
			lattice clocks beyond the {{Dick}} limit},\ }\href
	{https://doi.org/10.1038/nphoton.2011.34} {\bibfield  {journal} {\bibinfo
			{journal} {Nature Photonics}\ }\textbf {\bibinfo {volume} {5}},\ \bibinfo
		{pages} {288} (\bibinfo {year} {2011})}\BibitemShut {NoStop}%
	\bibitem [{\citenamefont {Li}\ \emph {et~al.}(2024)\citenamefont {Li},
		\citenamefont {Cui}, \citenamefont {Jia}, \citenamefont {Kong}, \citenamefont
		{Yu}, \citenamefont {Zhu}, \citenamefont {Liu}, \citenamefont {Wang},
		\citenamefont {Zhang}, \citenamefont {Huang}, \citenamefont {Zhu},
		\citenamefont {Yang}, \citenamefont {Hu}, \citenamefont {Liu}, \citenamefont
		{Zhai}, \citenamefont {Liu}, \citenamefont {Jiang}, \citenamefont {Xu},
		\citenamefont {Dai}, \citenamefont {Chen},\ and\ \citenamefont
		{Pan}}]{li2024strontium}%
	\BibitemOpen
	\bibfield  {author} {\bibinfo {author} {\bibfnamefont {J.}~\bibnamefont
			{Li}}, \bibinfo {author} {\bibfnamefont {X.-Y.}\ \bibnamefont {Cui}},
		\bibinfo {author} {\bibfnamefont {Z.-P.}\ \bibnamefont {Jia}}, \bibinfo
		{author} {\bibfnamefont {D.-Q.}\ \bibnamefont {Kong}}, \bibinfo {author}
		{\bibfnamefont {H.-W.}\ \bibnamefont {Yu}}, \bibinfo {author} {\bibfnamefont
			{X.-Q.}\ \bibnamefont {Zhu}}, \bibinfo {author} {\bibfnamefont {X.-Y.}\
			\bibnamefont {Liu}}, \bibinfo {author} {\bibfnamefont {D.-Z.}\ \bibnamefont
			{Wang}}, \bibinfo {author} {\bibfnamefont {X.}~\bibnamefont {Zhang}},
		\bibinfo {author} {\bibfnamefont {X.-Y.}\ \bibnamefont {Huang}}, \bibinfo
		{author} {\bibfnamefont {M.-Y.}\ \bibnamefont {Zhu}}, \bibinfo {author}
		{\bibfnamefont {Y.-M.}\ \bibnamefont {Yang}}, \bibinfo {author}
		{\bibfnamefont {Y.}~\bibnamefont {Hu}}, \bibinfo {author} {\bibfnamefont
			{X.-P.}\ \bibnamefont {Liu}}, \bibinfo {author} {\bibfnamefont {X.-M.}\
			\bibnamefont {Zhai}}, \bibinfo {author} {\bibfnamefont {P.}~\bibnamefont
			{Liu}}, \bibinfo {author} {\bibfnamefont {X.}~\bibnamefont {Jiang}}, \bibinfo
		{author} {\bibfnamefont {P.}~\bibnamefont {Xu}}, \bibinfo {author}
		{\bibfnamefont {H.-N.}\ \bibnamefont {Dai}}, \bibinfo {author} {\bibfnamefont
			{Y.-A.}\ \bibnamefont {Chen}},\ and\ \bibinfo {author} {\bibfnamefont
			{J.-W.}\ \bibnamefont {Pan}},\ }\bibfield  {title} {\bibinfo {title} {A
			strontium lattice clock with both stability and uncertainty below $5\times
			10^{-18}$},\ }\href {https://doi.org/10.1088/1681-7575/ad1a4c} {\bibfield
		{journal} {\bibinfo  {journal} {Metrologia}\ }\textbf {\bibinfo {volume}
			{61}},\ \bibinfo {pages} {015006} (\bibinfo {year} {2024})}\BibitemShut
	{NoStop}%
	\bibitem [{\citenamefont {Li}\ \emph {et~al.}(2023)\citenamefont {Li},
		\citenamefont {Jia}, \citenamefont {Liu}, \citenamefont {Liu}, \citenamefont
		{Wang}, \citenamefont {Kong}, \citenamefont {Li}, \citenamefont {Cui},
		\citenamefont {Dai}, \citenamefont {Chen},\ and\ \citenamefont
		{Pan}}]{Li2023}%
	\BibitemOpen
	\bibfield  {author} {\bibinfo {author} {\bibfnamefont {J.}~\bibnamefont
			{Li}}, \bibinfo {author} {\bibfnamefont {Z.-P.}\ \bibnamefont {Jia}},
		\bibinfo {author} {\bibfnamefont {P.}~\bibnamefont {Liu}}, \bibinfo {author}
		{\bibfnamefont {X.-Y.}\ \bibnamefont {Liu}}, \bibinfo {author} {\bibfnamefont
			{D.-Z.}\ \bibnamefont {Wang}}, \bibinfo {author} {\bibfnamefont {D.-Q.}\
			\bibnamefont {Kong}}, \bibinfo {author} {\bibfnamefont {S.-P.}\ \bibnamefont
			{Li}}, \bibinfo {author} {\bibfnamefont {X.-Y.}\ \bibnamefont {Cui}},
		\bibinfo {author} {\bibfnamefont {H.-N.}\ \bibnamefont {Dai}}, \bibinfo
		{author} {\bibfnamefont {Y.-A.}\ \bibnamefont {Chen}},\ and\ \bibinfo
		{author} {\bibfnamefont {J.-W.}\ \bibnamefont {Pan}},\ }\bibfield  {title}
	{\bibinfo {title} {An integrated high-flux cold atomic beam source for
			strontium},\ }\href {https://doi.org/10.1063/5.0162128} {\bibfield  {journal}
		{\bibinfo  {journal} {Review of Scientific Instruments}\ }\textbf {\bibinfo
			{volume} {94}},\ \bibinfo {pages} {093202} (\bibinfo {year}
		{2023})}\BibitemShut {NoStop}%
	\bibitem [{\citenamefont {Yu}\ \emph {et~al.}(2025)\citenamefont {Yu},
		\citenamefont {Liu}, \citenamefont {Li}, \citenamefont {Jia}, \citenamefont
		{Zhang}, \citenamefont {Yan}, \citenamefont {Li}, \citenamefont {Dai},\ and\
		\citenamefont {Chen}}]{yu2025}%
	\BibitemOpen
	\bibfield  {author} {\bibinfo {author} {\bibfnamefont {H.}~\bibnamefont
			{Yu}}, \bibinfo {author} {\bibfnamefont {P.}~\bibnamefont {Liu}}, \bibinfo
		{author} {\bibfnamefont {Y.}~\bibnamefont {Li}}, \bibinfo {author}
		{\bibfnamefont {Z.}~\bibnamefont {Jia}}, \bibinfo {author} {\bibfnamefont
			{X.}~\bibnamefont {Zhang}}, \bibinfo {author} {\bibfnamefont
			{J.}~\bibnamefont {Yan}}, \bibinfo {author} {\bibfnamefont {J.}~\bibnamefont
			{Li}}, \bibinfo {author} {\bibfnamefont {H.}~\bibnamefont {Dai}},\ and\
		\bibinfo {author} {\bibfnamefont {Y.}~\bibnamefont {Chen}},\ }\bibfield
	{title} {\bibinfo {title} {Improved evaluation of blackbody radiation shift
			with uncertainty below $1\times10^{-18}$ in a strontium lattice clock},\
	}\href {https://doi.org/10.1016/j.measurement.2025.118527} {\bibfield
		{journal} {\bibinfo  {journal} {Measurement}\ }\textbf {\bibinfo {volume}
			{257}},\ \bibinfo {pages} {118527} (\bibinfo {year} {2025})}\BibitemShut
	{NoStop}%
	\bibitem [{\citenamefont {Jia}\ \emph {et~al.}(2025)\citenamefont {Jia},
		\citenamefont {Li}, \citenamefont {Kong}, \citenamefont {Zhang},
		\citenamefont {Yu}, \citenamefont {Liu}, \citenamefont {Zhang}, \citenamefont
		{Wang}, \citenamefont {Zhu}, \citenamefont {Zhang}, \citenamefont {Zhu},
		\citenamefont {Feng}, \citenamefont {Cui}, \citenamefont {Xu}, \citenamefont
		{Jiang}, \citenamefont {Liu}, \citenamefont {Liu}, \citenamefont {Dai},
		\citenamefont {Chen},\ and\ \citenamefont {Pan}}]{jia2025}%
	\BibitemOpen
	\bibfield  {author} {\bibinfo {author} {\bibfnamefont {Z.-P.}\ \bibnamefont
			{Jia}}, \bibinfo {author} {\bibfnamefont {J.}~\bibnamefont {Li}}, \bibinfo
		{author} {\bibfnamefont {D.-Q.}\ \bibnamefont {Kong}}, \bibinfo {author}
		{\bibfnamefont {X.}~\bibnamefont {Zhang}}, \bibinfo {author} {\bibfnamefont
			{H.-W.}\ \bibnamefont {Yu}}, \bibinfo {author} {\bibfnamefont {X.-Y.}\
			\bibnamefont {Liu}}, \bibinfo {author} {\bibfnamefont {Y.-C.}\ \bibnamefont
			{Zhang}}, \bibinfo {author} {\bibfnamefont {Y.-B.}\ \bibnamefont {Wang}},
		\bibinfo {author} {\bibfnamefont {X.-Q.}\ \bibnamefont {Zhu}}, \bibinfo
		{author} {\bibfnamefont {J.-H.}\ \bibnamefont {Zhang}}, \bibinfo {author}
		{\bibfnamefont {M.-Y.}\ \bibnamefont {Zhu}}, \bibinfo {author} {\bibfnamefont
			{P.-J.}\ \bibnamefont {Feng}}, \bibinfo {author} {\bibfnamefont {X.-Y.}\
			\bibnamefont {Cui}}, \bibinfo {author} {\bibfnamefont {P.}~\bibnamefont
			{Xu}}, \bibinfo {author} {\bibfnamefont {X.}~\bibnamefont {Jiang}}, \bibinfo
		{author} {\bibfnamefont {X.-P.}\ \bibnamefont {Liu}}, \bibinfo {author}
		{\bibfnamefont {P.}~\bibnamefont {Liu}}, \bibinfo {author} {\bibfnamefont
			{H.-N.}\ \bibnamefont {Dai}}, \bibinfo {author} {\bibfnamefont {Y.-A.}\
			\bibnamefont {Chen}},\ and\ \bibinfo {author} {\bibfnamefont {J.-W.}\
			\bibnamefont {Pan}},\ }\href {https://arxiv.org/abs/2509.13991} {\bibinfo
		{title} {Improved systematic evaluation of a strontium optical clock with
			uncertainty below $1\times 10^{-18}$}} (\bibinfo {year} {2025}),\ \Eprint
	{https://arxiv.org/abs/2509.13991} {arXiv:2509.13991 [physics.atom-ph]}
	\BibitemShut {NoStop}%
	\bibitem [{sup()}]{supp}%
	\BibitemOpen
	\href@noop {} {\bibinfo {title} {{Supplemental Material}}},\ \bibinfo {note}
	{includes experimental setup, performance of OLOs, Dick effect calculation,
		simulation of clock comparison, and clock comparison results at different
		times.}\BibitemShut {Stop}%
	\bibitem [{\citenamefont {Zhu}\ \emph {et~al.}(2024)\citenamefont {Zhu},
		\citenamefont {Cui}, \citenamefont {Kong}, \citenamefont {Yu}, \citenamefont
		{Zhai}, \citenamefont {Zheng}, \citenamefont {Xie}, \citenamefont {Zhang},
		\citenamefont {Jiang}, \citenamefont {Zhang}, \citenamefont {Xu},
		\citenamefont {Dai}, \citenamefont {Chen},\ and\ \citenamefont
		{Pan}}]{zhu2024}%
	\BibitemOpen
	\bibfield  {author} {\bibinfo {author} {\bibfnamefont {X.-Q.}\ \bibnamefont
			{Zhu}}, \bibinfo {author} {\bibfnamefont {X.-Y.}\ \bibnamefont {Cui}},
		\bibinfo {author} {\bibfnamefont {D.-Q.}\ \bibnamefont {Kong}}, \bibinfo
		{author} {\bibfnamefont {H.-W.}\ \bibnamefont {Yu}}, \bibinfo {author}
		{\bibfnamefont {X.-M.}\ \bibnamefont {Zhai}}, \bibinfo {author}
		{\bibfnamefont {M.-Y.}\ \bibnamefont {Zheng}}, \bibinfo {author}
		{\bibfnamefont {X.-P.}\ \bibnamefont {Xie}}, \bibinfo {author} {\bibfnamefont
			{Q.}~\bibnamefont {Zhang}}, \bibinfo {author} {\bibfnamefont
			{X.}~\bibnamefont {Jiang}}, \bibinfo {author} {\bibfnamefont {X.-B.}\
			\bibnamefont {Zhang}}, \bibinfo {author} {\bibfnamefont {P.}~\bibnamefont
			{Xu}}, \bibinfo {author} {\bibfnamefont {H.-N.}\ \bibnamefont {Dai}},
		\bibinfo {author} {\bibfnamefont {Y.-A.}\ \bibnamefont {Chen}},\ and\
		\bibinfo {author} {\bibfnamefont {J.-W.}\ \bibnamefont {Pan}},\ }\bibfield
	{title} {\bibinfo {title} {An ultrastable 1397-nm laser stabilized by a
			crystalline-coated room-temperature cavity},\ }\href
	{https://doi.org/10.1063/5.0200553} {\bibfield  {journal} {\bibinfo
			{journal} {Review of Scientific Instruments}\ }\textbf {\bibinfo {volume}
			{95}},\ \bibinfo {pages} {083002} (\bibinfo {year} {2024})}\BibitemShut
	{NoStop}%
	\bibitem [{\citenamefont {Ohmae}\ \emph {et~al.}(2021)\citenamefont {Ohmae},
		\citenamefont {Takamoto}, \citenamefont {Takahashi}, \citenamefont {Kokubun},
		\citenamefont {Araki}, \citenamefont {Hinton}, \citenamefont {Ushijima},
		\citenamefont {Muramatsu}, \citenamefont {Furumiya}, \citenamefont {Sakai},
		\citenamefont {Moriya}, \citenamefont {Kamiya}, \citenamefont {Fujii},
		\citenamefont {Muramatsu}, \citenamefont {Shiimado},\ and\ \citenamefont
		{Katori}}]{ohmae2021}%
	\BibitemOpen
	\bibfield  {author} {\bibinfo {author} {\bibfnamefont {N.}~\bibnamefont
			{Ohmae}}, \bibinfo {author} {\bibfnamefont {M.}~\bibnamefont {Takamoto}},
		\bibinfo {author} {\bibfnamefont {Y.}~\bibnamefont {Takahashi}}, \bibinfo
		{author} {\bibfnamefont {M.}~\bibnamefont {Kokubun}}, \bibinfo {author}
		{\bibfnamefont {K.}~\bibnamefont {Araki}}, \bibinfo {author} {\bibfnamefont
			{A.}~\bibnamefont {Hinton}}, \bibinfo {author} {\bibfnamefont
			{I.}~\bibnamefont {Ushijima}}, \bibinfo {author} {\bibfnamefont
			{T.}~\bibnamefont {Muramatsu}}, \bibinfo {author} {\bibfnamefont
			{T.}~\bibnamefont {Furumiya}}, \bibinfo {author} {\bibfnamefont
			{Y.}~\bibnamefont {Sakai}}, \bibinfo {author} {\bibfnamefont
			{N.}~\bibnamefont {Moriya}}, \bibinfo {author} {\bibfnamefont
			{N.}~\bibnamefont {Kamiya}}, \bibinfo {author} {\bibfnamefont
			{K.}~\bibnamefont {Fujii}}, \bibinfo {author} {\bibfnamefont
			{R.}~\bibnamefont {Muramatsu}}, \bibinfo {author} {\bibfnamefont
			{T.}~\bibnamefont {Shiimado}},\ and\ \bibinfo {author} {\bibfnamefont
			{H.}~\bibnamefont {Katori}},\ }\bibfield  {title} {\bibinfo {title}
		{Transportable {{Strontium Optical Lattice Clocks Operated Outside
					Laboratory}} at the {{Level}} of $10^{-18}$ {{Uncertainty}}},\ }\href
	{https://doi.org/10.1002/qute.202100015} {\bibfield  {journal} {\bibinfo
			{journal} {Advanced Quantum Technologies}\ }\textbf {\bibinfo {volume} {4}},\
		\bibinfo {pages} {2100015} (\bibinfo {year} {2021})}\BibitemShut {NoStop}%
	\bibitem [{\citenamefont {Bothwell}\ \emph {et~al.}(2025)\citenamefont
		{Bothwell}, \citenamefont {Brand}, \citenamefont {Fasano}, \citenamefont
		{Akin}, \citenamefont {Whalen}, \citenamefont {Grogan}, \citenamefont {Chen},
		\citenamefont {Pomponio}, \citenamefont {Nakamura}, \citenamefont {Rauf},
		\citenamefont {Baldoni}, \citenamefont {Giunta}, \citenamefont {Holzwarth},
		\citenamefont {Nelson}, \citenamefont {Hati}, \citenamefont {Quinlan},
		\citenamefont {Fox}, \citenamefont {Peil},\ and\ \citenamefont
		{Ludlow}}]{bothwell2025}%
	\BibitemOpen
	\bibfield  {author} {\bibinfo {author} {\bibfnamefont {T.}~\bibnamefont
			{Bothwell}}, \bibinfo {author} {\bibfnamefont {W.}~\bibnamefont {Brand}},
		\bibinfo {author} {\bibfnamefont {R.}~\bibnamefont {Fasano}}, \bibinfo
		{author} {\bibfnamefont {T.}~\bibnamefont {Akin}}, \bibinfo {author}
		{\bibfnamefont {J.}~\bibnamefont {Whalen}}, \bibinfo {author} {\bibfnamefont
			{T.}~\bibnamefont {Grogan}}, \bibinfo {author} {\bibfnamefont {Y.-J.}\
			\bibnamefont {Chen}}, \bibinfo {author} {\bibfnamefont {M.}~\bibnamefont
			{Pomponio}}, \bibinfo {author} {\bibfnamefont {T.}~\bibnamefont {Nakamura}},
		\bibinfo {author} {\bibfnamefont {B.}~\bibnamefont {Rauf}}, \bibinfo {author}
		{\bibfnamefont {I.}~\bibnamefont {Baldoni}}, \bibinfo {author} {\bibfnamefont
			{M.}~\bibnamefont {Giunta}}, \bibinfo {author} {\bibfnamefont
			{R.}~\bibnamefont {Holzwarth}}, \bibinfo {author} {\bibfnamefont
			{C.}~\bibnamefont {Nelson}}, \bibinfo {author} {\bibfnamefont
			{A.}~\bibnamefont {Hati}}, \bibinfo {author} {\bibfnamefont {F.}~\bibnamefont
			{Quinlan}}, \bibinfo {author} {\bibfnamefont {R.}~\bibnamefont {Fox}},
		\bibinfo {author} {\bibfnamefont {S.}~\bibnamefont {Peil}},\ and\ \bibinfo
		{author} {\bibfnamefont {A.}~\bibnamefont {Ludlow}},\ }\bibfield  {title}
	{\bibinfo {title} {Deployment of a transportable {{Yb}} optical lattice
			clock},\ }\href {https://doi.org/10.1364/OL.543310} {\bibfield  {journal}
		{\bibinfo  {journal} {Optics Letters}\ }\textbf {\bibinfo {volume} {50}},\
		\bibinfo {pages} {646} (\bibinfo {year} {2025})}\BibitemShut {NoStop}%
	\bibitem [{\citenamefont {Bongs}\ \emph {et~al.}(2015)\citenamefont {Bongs},
		\citenamefont {Singh}, \citenamefont {Smith}, \citenamefont {He},
		\citenamefont {Kock}, \citenamefont {Swierad}, \citenamefont {Hughes},
		\citenamefont {Schiller}, \citenamefont {Alighanbari}, \citenamefont
		{Origlia}, \citenamefont {Vogt}, \citenamefont {Sterr}, \citenamefont
		{Lisdat}, \citenamefont {Targat}, \citenamefont {Lodewyck}, \citenamefont
		{Holleville}, \citenamefont {Venon}, \citenamefont {Bize}, \citenamefont
		{Barwood}, \citenamefont {Gill}, \citenamefont {Hill}, \citenamefont
		{Ovchinnikov}, \citenamefont {Poli}, \citenamefont {Tino}, \citenamefont
		{Stuhler}, \citenamefont {Kaenders},\ and\ \citenamefont
		{{team}}}]{bongs2015}%
	\BibitemOpen
	\bibfield  {author} {\bibinfo {author} {\bibfnamefont {K.}~\bibnamefont
			{Bongs}}, \bibinfo {author} {\bibfnamefont {Y.}~\bibnamefont {Singh}},
		\bibinfo {author} {\bibfnamefont {L.}~\bibnamefont {Smith}}, \bibinfo
		{author} {\bibfnamefont {W.}~\bibnamefont {He}}, \bibinfo {author}
		{\bibfnamefont {O.}~\bibnamefont {Kock}}, \bibinfo {author} {\bibfnamefont
			{D.}~\bibnamefont {Swierad}}, \bibinfo {author} {\bibfnamefont
			{J.}~\bibnamefont {Hughes}}, \bibinfo {author} {\bibfnamefont
			{S.}~\bibnamefont {Schiller}}, \bibinfo {author} {\bibfnamefont
			{S.}~\bibnamefont {Alighanbari}}, \bibinfo {author} {\bibfnamefont
			{S.}~\bibnamefont {Origlia}}, \bibinfo {author} {\bibfnamefont
			{S.}~\bibnamefont {Vogt}}, \bibinfo {author} {\bibfnamefont {U.}~\bibnamefont
			{Sterr}}, \bibinfo {author} {\bibfnamefont {C.}~\bibnamefont {Lisdat}},
		\bibinfo {author} {\bibfnamefont {R.~L.}\ \bibnamefont {Targat}}, \bibinfo
		{author} {\bibfnamefont {J.}~\bibnamefont {Lodewyck}}, \bibinfo {author}
		{\bibfnamefont {D.}~\bibnamefont {Holleville}}, \bibinfo {author}
		{\bibfnamefont {B.}~\bibnamefont {Venon}}, \bibinfo {author} {\bibfnamefont
			{S.}~\bibnamefont {Bize}}, \bibinfo {author} {\bibfnamefont {G.~P.}\
			\bibnamefont {Barwood}}, \bibinfo {author} {\bibfnamefont {P.}~\bibnamefont
			{Gill}}, \bibinfo {author} {\bibfnamefont {I.~R.}\ \bibnamefont {Hill}},
		\bibinfo {author} {\bibfnamefont {Y.~B.}\ \bibnamefont {Ovchinnikov}},
		\bibinfo {author} {\bibfnamefont {N.}~\bibnamefont {Poli}}, \bibinfo {author}
		{\bibfnamefont {G.~M.}\ \bibnamefont {Tino}}, \bibinfo {author}
		{\bibfnamefont {J.}~\bibnamefont {Stuhler}}, \bibinfo {author} {\bibfnamefont
			{W.}~\bibnamefont {Kaenders}},\ and\ \bibinfo {author} {\bibfnamefont
			{t.~S.}\ \bibnamefont {{team}}},\ }\bibfield  {title} {\bibinfo {title}
		{Development of a strontium optical lattice clock for the {{SOC}} mission on
			the {{ISS}}},\ }\href {https://doi.org/10.1016/j.crhy.2015.03.009} {\bibfield
		{journal} {\bibinfo  {journal} {Comptes Rendus. Physique}\ }\textbf
		{\bibinfo {volume} {16}},\ \bibinfo {pages} {553} (\bibinfo {year}
		{2015})}\BibitemShut {NoStop}%
	\bibitem [{\citenamefont {Derevianko}\ \emph {et~al.}(2022)\citenamefont
		{Derevianko}, \citenamefont {Gibble}, \citenamefont {Hollberg}, \citenamefont
		{Newbury}, \citenamefont {Oates}, \citenamefont {Safronova}, \citenamefont
		{Sinclair},\ and\ \citenamefont {Yu}}]{derevianko2022}%
	\BibitemOpen
	\bibfield  {author} {\bibinfo {author} {\bibfnamefont {A.}~\bibnamefont
			{Derevianko}}, \bibinfo {author} {\bibfnamefont {K.}~\bibnamefont {Gibble}},
		\bibinfo {author} {\bibfnamefont {L.}~\bibnamefont {Hollberg}}, \bibinfo
		{author} {\bibfnamefont {N.~R.}\ \bibnamefont {Newbury}}, \bibinfo {author}
		{\bibfnamefont {C.}~\bibnamefont {Oates}}, \bibinfo {author} {\bibfnamefont
			{M.~S.}\ \bibnamefont {Safronova}}, \bibinfo {author} {\bibfnamefont {L.~C.}\
			\bibnamefont {Sinclair}},\ and\ \bibinfo {author} {\bibfnamefont
			{N.}~\bibnamefont {Yu}},\ }\bibfield  {title} {\bibinfo {title} {Fundamental
			physics with a state-of-the-art optical clock in space},\ }\href
	{https://doi.org/10.1088/2058-9565/ac7df9} {\bibfield  {journal} {\bibinfo
			{journal} {Quantum Science and Technology}\ }\textbf {\bibinfo {volume}
			{7}},\ \bibinfo {pages} {044002} (\bibinfo {year} {2022})}\BibitemShut
	{NoStop}%
	\bibitem [{\citenamefont {Schkolnik}\ \emph {et~al.}(2023)\citenamefont
		{Schkolnik}, \citenamefont {Budker}, \citenamefont {Fartmann}, \citenamefont
		{Flambaum}, \citenamefont {Hollberg}, \citenamefont {Kalaydzhyan},
		\citenamefont {Kolkowitz}, \citenamefont {Krutzik}, \citenamefont {Ludlow},
		\citenamefont {Newbury}, \citenamefont {Pyrlik}, \citenamefont {Sinclair},
		\citenamefont {Stadnik}, \citenamefont {Tietje}, \citenamefont {Ye},\ and\
		\citenamefont {Williams}}]{schkolnik2023}%
	\BibitemOpen
	\bibfield  {author} {\bibinfo {author} {\bibfnamefont {V.}~\bibnamefont
			{Schkolnik}}, \bibinfo {author} {\bibfnamefont {D.}~\bibnamefont {Budker}},
		\bibinfo {author} {\bibfnamefont {O.}~\bibnamefont {Fartmann}}, \bibinfo
		{author} {\bibfnamefont {V.}~\bibnamefont {Flambaum}}, \bibinfo {author}
		{\bibfnamefont {L.}~\bibnamefont {Hollberg}}, \bibinfo {author}
		{\bibfnamefont {T.}~\bibnamefont {Kalaydzhyan}}, \bibinfo {author}
		{\bibfnamefont {S.}~\bibnamefont {Kolkowitz}}, \bibinfo {author}
		{\bibfnamefont {M.}~\bibnamefont {Krutzik}}, \bibinfo {author} {\bibfnamefont
			{A.}~\bibnamefont {Ludlow}}, \bibinfo {author} {\bibfnamefont
			{N.}~\bibnamefont {Newbury}}, \bibinfo {author} {\bibfnamefont
			{C.}~\bibnamefont {Pyrlik}}, \bibinfo {author} {\bibfnamefont
			{L.}~\bibnamefont {Sinclair}}, \bibinfo {author} {\bibfnamefont
			{Y.}~\bibnamefont {Stadnik}}, \bibinfo {author} {\bibfnamefont
			{I.}~\bibnamefont {Tietje}}, \bibinfo {author} {\bibfnamefont
			{J.}~\bibnamefont {Ye}},\ and\ \bibinfo {author} {\bibfnamefont
			{J.}~\bibnamefont {Williams}},\ }\bibfield  {title} {\bibinfo {title}
		{Optical atomic clock aboard an {{Earth-orbiting}} space station
			({{OACESS}}): Enhancing searches for physics beyond the standard model in
			space},\ }\href {https://doi.org/10.1088/2058-9565/ac9f2b} {\bibfield
		{journal} {\bibinfo  {journal} {Quantum Science and Technology}\ }\textbf
		{\bibinfo {volume} {8}},\ \bibinfo {pages} {014003} (\bibinfo {year}
		{2023})}\BibitemShut {NoStop}%
\end{thebibliography}

\begin{thebibliography}{10}%
	\makeatletter
	\providecommand \@ifxundefined [1]{%
		\@ifx{#1\undefined}
	}%
	\providecommand \@ifnum [1]{%
		\ifnum #1\expandafter \@firstoftwo
		\else \expandafter \@secondoftwo
		\fi
	}%
	\providecommand \@ifx [1]{%
		\ifx #1\expandafter \@firstoftwo
		\else \expandafter \@secondoftwo
		\fi
	}%
	\providecommand \natexlab [1]{#1}%
	\providecommand \enquote  [1]{``#1''}%
	\providecommand \bibnamefont  [1]{#1}%
	\providecommand \bibfnamefont [1]{#1}%
	\providecommand \citenamefont [1]{#1}%
	\providecommand \href@noop [0]{\@secondoftwo}%
	\providecommand \href [0]{\begingroup \@sanitize@url \@href}%
	\providecommand \@href[1]{\@@startlink{#1}\@@href}%
	\providecommand \@@href[1]{\endgroup#1\@@endlink}%
	\providecommand \@sanitize@url [0]{\catcode `\\12\catcode `\$12\catcode
		`\&12\catcode `\#12\catcode `\^12\catcode `\_12\catcode `\%12\relax}%
	\providecommand \@@startlink[1]{}%
	\providecommand \@@endlink[0]{}%
	\providecommand \url  [0]{\begingroup\@sanitize@url \@url }%
	\providecommand \@url [1]{\endgroup\@href {#1}{\urlprefix }}%
	\providecommand \urlprefix  [0]{URL }%
	\providecommand \Eprint [0]{\href }%
	\providecommand \doibase [0]{https://doi.org/}%
	\providecommand \selectlanguage [0]{\@gobble}%
	\providecommand \bibinfo  [0]{\@secondoftwo}%
	\providecommand \bibfield  [0]{\@secondoftwo}%
	\providecommand \translation [1]{[#1]}%
	\providecommand \BibitemOpen [0]{}%
	\providecommand \bibitemStop [0]{}%
	\providecommand \bibitemNoStop [0]{.\EOS\space}%
	\providecommand \EOS [0]{\spacefactor3000\relax}%
	\providecommand \BibitemShut  [1]{\csname bibitem#1\endcsname}%
	\let\auto@bib@innerbib\@empty
	%</preamble>
	\bibitem [{\citenamefont {Li}\ \emph {et~al.}(2024)\citenamefont {Li},
		\citenamefont {Cui}, \citenamefont {Jia}, \citenamefont {Kong}, \citenamefont
		{Yu}, \citenamefont {Zhu}, \citenamefont {Liu}, \citenamefont {Wang},
		\citenamefont {Zhang}, \citenamefont {Huang}, \citenamefont {Zhu},
		\citenamefont {Yang}, \citenamefont {Hu}, \citenamefont {Liu}, \citenamefont
		{Zhai}, \citenamefont {Liu}, \citenamefont {Jiang}, \citenamefont {Xu},
		\citenamefont {Dai}, \citenamefont {Chen},\ and\ \citenamefont
		{Pan}}]{li2024strontium}%
	\BibitemOpen
	\bibfield  {author} {\bibinfo {author} {\bibfnamefont {J.}~\bibnamefont
			{Li}}, \bibinfo {author} {\bibfnamefont {X.-Y.}\ \bibnamefont {Cui}},
		\bibinfo {author} {\bibfnamefont {Z.-P.}\ \bibnamefont {Jia}}, \bibinfo
		{author} {\bibfnamefont {D.-Q.}\ \bibnamefont {Kong}}, \bibinfo {author}
		{\bibfnamefont {H.-W.}\ \bibnamefont {Yu}}, \bibinfo {author} {\bibfnamefont
			{X.-Q.}\ \bibnamefont {Zhu}}, \bibinfo {author} {\bibfnamefont {X.-Y.}\
			\bibnamefont {Liu}}, \bibinfo {author} {\bibfnamefont {D.-Z.}\ \bibnamefont
			{Wang}}, \bibinfo {author} {\bibfnamefont {X.}~\bibnamefont {Zhang}},
		\bibinfo {author} {\bibfnamefont {X.-Y.}\ \bibnamefont {Huang}}, \bibinfo
		{author} {\bibfnamefont {M.-Y.}\ \bibnamefont {Zhu}}, \bibinfo {author}
		{\bibfnamefont {Y.-M.}\ \bibnamefont {Yang}}, \bibinfo {author}
		{\bibfnamefont {Y.}~\bibnamefont {Hu}}, \bibinfo {author} {\bibfnamefont
			{X.-P.}\ \bibnamefont {Liu}}, \bibinfo {author} {\bibfnamefont {X.-M.}\
			\bibnamefont {Zhai}}, \bibinfo {author} {\bibfnamefont {P.}~\bibnamefont
			{Liu}}, \bibinfo {author} {\bibfnamefont {X.}~\bibnamefont {Jiang}}, \bibinfo
		{author} {\bibfnamefont {P.}~\bibnamefont {Xu}}, \bibinfo {author}
		{\bibfnamefont {H.-N.}\ \bibnamefont {Dai}}, \bibinfo {author} {\bibfnamefont
			{Y.-A.}\ \bibnamefont {Chen}},\ and\ \bibinfo {author} {\bibfnamefont
			{J.-W.}\ \bibnamefont {Pan}},\ }\href
	{https://doi.org/10.1088/1681-7575/ad1a4c} {\bibfield  {journal} {\bibinfo
			{journal} {Metrologia}\ }\textbf {\bibinfo {volume} {61}},\ \bibinfo {pages}
		{015006} (\bibinfo {year} {2024})}\BibitemShut {NoStop}%
	\bibitem [{\citenamefont {Li}\ \emph {et~al.}(2023)\citenamefont {Li},
		\citenamefont {Jia}, \citenamefont {Liu}, \citenamefont {Liu}, \citenamefont
		{Wang}, \citenamefont {Kong}, \citenamefont {Li}, \citenamefont {Cui},
		\citenamefont {Dai}, \citenamefont {Chen},\ and\ \citenamefont
		{Pan}}]{Li2023}%
	\BibitemOpen
	\bibfield  {author} {\bibinfo {author} {\bibfnamefont {J.}~\bibnamefont
			{Li}}, \bibinfo {author} {\bibfnamefont {Z.-P.}\ \bibnamefont {Jia}},
		\bibinfo {author} {\bibfnamefont {P.}~\bibnamefont {Liu}}, \bibinfo {author}
		{\bibfnamefont {X.-Y.}\ \bibnamefont {Liu}}, \bibinfo {author} {\bibfnamefont
			{D.-Z.}\ \bibnamefont {Wang}}, \bibinfo {author} {\bibfnamefont {D.-Q.}\
			\bibnamefont {Kong}}, \bibinfo {author} {\bibfnamefont {S.-P.}\ \bibnamefont
			{Li}}, \bibinfo {author} {\bibfnamefont {X.-Y.}\ \bibnamefont {Cui}},
		\bibinfo {author} {\bibfnamefont {H.-N.}\ \bibnamefont {Dai}}, \bibinfo
		{author} {\bibfnamefont {Y.-A.}\ \bibnamefont {Chen}},\ and\ \bibinfo
		{author} {\bibfnamefont {J.-W.}\ \bibnamefont {Pan}},\ }\href
	{https://doi.org/10.1063/5.0162128} {\bibfield  {journal} {\bibinfo
			{journal} {Review of Scientific Instruments}\ }\textbf {\bibinfo {volume}
			{94}},\ \bibinfo {pages} {093202} (\bibinfo {year} {2023})}\BibitemShut
	{NoStop}%
	\bibitem [{\citenamefont {Jia}\ \emph {et~al.}(2025{\natexlab{a}})\citenamefont
		{Jia}, \citenamefont {Li}, \citenamefont {Kong}, \citenamefont {Zhang},
		\citenamefont {Yu}, \citenamefont {Liu}, \citenamefont {Zhang}, \citenamefont
		{Wang}, \citenamefont {Zhu}, \citenamefont {Zhang}, \citenamefont {Zhu},
		\citenamefont {Feng}, \citenamefont {Cui}, \citenamefont {Xu}, \citenamefont
		{Jiang}, \citenamefont {Liu}, \citenamefont {Liu}, \citenamefont {Dai},
		\citenamefont {Chen},\ and\ \citenamefont {Pan}}]{jia2025uncertainty}%
	\BibitemOpen
	\bibfield  {author} {\bibinfo {author} {\bibfnamefont {Z.-P.}\ \bibnamefont
			{Jia}}, \bibinfo {author} {\bibfnamefont {J.}~\bibnamefont {Li}}, \bibinfo
		{author} {\bibfnamefont {D.-Q.}\ \bibnamefont {Kong}}, \bibinfo {author}
		{\bibfnamefont {X.}~\bibnamefont {Zhang}}, \bibinfo {author} {\bibfnamefont
			{H.-W.}\ \bibnamefont {Yu}}, \bibinfo {author} {\bibfnamefont {X.-Y.}\
			\bibnamefont {Liu}}, \bibinfo {author} {\bibfnamefont {Y.-C.}\ \bibnamefont
			{Zhang}}, \bibinfo {author} {\bibfnamefont {Y.-B.}\ \bibnamefont {Wang}},
		\bibinfo {author} {\bibfnamefont {X.-Q.}\ \bibnamefont {Zhu}}, \bibinfo
		{author} {\bibfnamefont {J.-H.}\ \bibnamefont {Zhang}}, \bibinfo {author}
		{\bibfnamefont {M.-Y.}\ \bibnamefont {Zhu}}, \bibinfo {author} {\bibfnamefont
			{P.-J.}\ \bibnamefont {Feng}}, \bibinfo {author} {\bibfnamefont {X.-Y.}\
			\bibnamefont {Cui}}, \bibinfo {author} {\bibfnamefont {P.}~\bibnamefont
			{Xu}}, \bibinfo {author} {\bibfnamefont {X.}~\bibnamefont {Jiang}}, \bibinfo
		{author} {\bibfnamefont {X.-P.}\ \bibnamefont {Liu}}, \bibinfo {author}
		{\bibfnamefont {P.}~\bibnamefont {Liu}}, \bibinfo {author} {\bibfnamefont
			{H.-N.}\ \bibnamefont {Dai}}, \bibinfo {author} {\bibfnamefont {Y.-A.}\
			\bibnamefont {Chen}},\ and\ \bibinfo {author} {\bibfnamefont {J.-W.}\
			\bibnamefont {Pan}},\ }\href {https://arxiv.org/abs/2509.13991} {\bibinfo
		{title} {Improved systematic evaluation of a strontium optical clock with
			uncertainty below $1\times 10^{-18}$}} (\bibinfo {year}
	{2025}{\natexlab{a}}),\ \Eprint {https://arxiv.org/abs/2509.13991}
	{arXiv:2509.13991 [physics.atom-ph]} \BibitemShut {NoStop}%
	\bibitem [{\citenamefont {Jia}\ \emph {et~al.}(2025{\natexlab{b}})\citenamefont
		{Jia}, \citenamefont {Cui}, \citenamefont {Xie}, \citenamefont {Zhang},
		\citenamefont {Niu}, \citenamefont {Liu}, \citenamefont {Zhu}, \citenamefont
		{Li},\ and\ \citenamefont {Dai}}]{Jia2025}%
	\BibitemOpen
	\bibfield  {author} {\bibinfo {author} {\bibfnamefont {Z.-P.}\ \bibnamefont
			{Jia}}, \bibinfo {author} {\bibfnamefont {X.-Y.}\ \bibnamefont {Cui}},
		\bibinfo {author} {\bibfnamefont {Y.-J.}\ \bibnamefont {Xie}}, \bibinfo
		{author} {\bibfnamefont {X.}~\bibnamefont {Zhang}}, \bibinfo {author}
		{\bibfnamefont {G.-Z.}\ \bibnamefont {Niu}}, \bibinfo {author} {\bibfnamefont
			{X.-Y.}\ \bibnamefont {Liu}}, \bibinfo {author} {\bibfnamefont {Q.-Q.}\
			\bibnamefont {Zhu}}, \bibinfo {author} {\bibfnamefont {J.}~\bibnamefont
			{Li}},\ and\ \bibinfo {author} {\bibfnamefont {H.-N.}\ \bibnamefont {Dai}},\
	}\bibfield  {journal} {\bibinfo  {journal} {Physical Review Applied}\
	}\textbf {\bibinfo {volume} {23}},\ \href
	{https://doi.org/10.1103/physrevapplied.23.014014}
	{10.1103/physrevapplied.23.014014} (\bibinfo {year}
	{2025}{\natexlab{b}})\BibitemShut {NoStop}%
	\bibitem [{\citenamefont {Fasano}\ \emph {et~al.}(2021)\citenamefont {Fasano},
		\citenamefont {Chen}, \citenamefont {McGrew}, \citenamefont {Brand},
		\citenamefont {Fox},\ and\ \citenamefont {Ludlow}}]{Fasano2021}%
	\BibitemOpen
	\bibfield  {author} {\bibinfo {author} {\bibfnamefont {R.}~\bibnamefont
			{Fasano}}, \bibinfo {author} {\bibfnamefont {Y.}~\bibnamefont {Chen}},
		\bibinfo {author} {\bibfnamefont {W.}~\bibnamefont {McGrew}}, \bibinfo
		{author} {\bibfnamefont {W.}~\bibnamefont {Brand}}, \bibinfo {author}
		{\bibfnamefont {R.}~\bibnamefont {Fox}},\ and\ \bibinfo {author}
		{\bibfnamefont {A.}~\bibnamefont {Ludlow}},\ }\href
	{https://doi.org/10.1103/PhysRevApplied.15.044016} {\bibfield  {journal}
		{\bibinfo  {journal} {Phys. Rev. Appl.}\ }\textbf {\bibinfo {volume} {15}},\
		\bibinfo {pages} {044016} (\bibinfo {year} {2021})}\BibitemShut {NoStop}%
	\bibitem [{\citenamefont {Oelker}\ \emph {et~al.}(2019)\citenamefont {Oelker},
		\citenamefont {Hutson}, \citenamefont {Kennedy}, \citenamefont {Sonderhouse},
		\citenamefont {Bothwell}, \citenamefont {Goban}, \citenamefont {Kedar},
		\citenamefont {Sanner}, \citenamefont {Robinson}, \citenamefont {Marti},
		\citenamefont {Matei}, \citenamefont {Legero}, \citenamefont {Giunta},
		\citenamefont {Holzwarth}, \citenamefont {Riehle}, \citenamefont {Sterr},\
		and\ \citenamefont {Ye}}]{oelker2019}%
	\BibitemOpen
	\bibfield  {author} {\bibinfo {author} {\bibfnamefont {E.}~\bibnamefont
			{Oelker}}, \bibinfo {author} {\bibfnamefont {R.~B.}\ \bibnamefont {Hutson}},
		\bibinfo {author} {\bibfnamefont {C.~J.}\ \bibnamefont {Kennedy}}, \bibinfo
		{author} {\bibfnamefont {L.}~\bibnamefont {Sonderhouse}}, \bibinfo {author}
		{\bibfnamefont {T.}~\bibnamefont {Bothwell}}, \bibinfo {author}
		{\bibfnamefont {A.}~\bibnamefont {Goban}}, \bibinfo {author} {\bibfnamefont
			{D.}~\bibnamefont {Kedar}}, \bibinfo {author} {\bibfnamefont
			{C.}~\bibnamefont {Sanner}}, \bibinfo {author} {\bibfnamefont {J.~M.}\
			\bibnamefont {Robinson}}, \bibinfo {author} {\bibfnamefont {G.~E.}\
			\bibnamefont {Marti}}, \bibinfo {author} {\bibfnamefont {D.~G.}\ \bibnamefont
			{Matei}}, \bibinfo {author} {\bibfnamefont {T.}~\bibnamefont {Legero}},
		\bibinfo {author} {\bibfnamefont {M.}~\bibnamefont {Giunta}}, \bibinfo
		{author} {\bibfnamefont {R.}~\bibnamefont {Holzwarth}}, \bibinfo {author}
		{\bibfnamefont {F.}~\bibnamefont {Riehle}}, \bibinfo {author} {\bibfnamefont
			{U.}~\bibnamefont {Sterr}},\ and\ \bibinfo {author} {\bibfnamefont
			{J.}~\bibnamefont {Ye}},\ }\href {https://doi.org/10.1038/s41566-019-0493-4}
	{\bibfield  {journal} {\bibinfo  {journal} {Nature Photonics}\ }\textbf
		{\bibinfo {volume} {13}},\ \bibinfo {pages} {714} (\bibinfo {year}
		{2019})}\BibitemShut {NoStop}%
	\bibitem [{\citenamefont {Ushijima}\ \emph {et~al.}(2018)\citenamefont
		{Ushijima}, \citenamefont {Takamoto},\ and\ \citenamefont
		{Katori}}]{Ushijima2018}%
	\BibitemOpen
	\bibfield  {author} {\bibinfo {author} {\bibfnamefont {I.}~\bibnamefont
			{Ushijima}}, \bibinfo {author} {\bibfnamefont {M.}~\bibnamefont {Takamoto}},\
		and\ \bibinfo {author} {\bibfnamefont {H.}~\bibnamefont {Katori}},\ }\href
	{https://doi.org/10.1103/PhysRevLett.121.263202} {\bibfield  {journal}
		{\bibinfo  {journal} {Phys. Rev. Lett.}\ }\textbf {\bibinfo {volume} {121}},\
		\bibinfo {pages} {263202} (\bibinfo {year} {2018})}\BibitemShut {NoStop}%
	\bibitem [{\citenamefont {Falke}\ \emph {et~al.}(2014)\citenamefont {Falke},
		\citenamefont {Lemke}, \citenamefont {Grebing}, \citenamefont {Lipphardt},
		\citenamefont {Weyers}, \citenamefont {Gerginov}, \citenamefont {Huntemann},
		\citenamefont {Hagemann}, \citenamefont {Al-Masoudi}, \citenamefont
		{H\"{a}fner}, \citenamefont {Vogt}, \citenamefont {Sterr},\ and\
		\citenamefont {Lisdat}}]{Falke2014}%
	\BibitemOpen
	\bibfield  {author} {\bibinfo {author} {\bibfnamefont {S.}~\bibnamefont
			{Falke}}, \bibinfo {author} {\bibfnamefont {N.}~\bibnamefont {Lemke}},
		\bibinfo {author} {\bibfnamefont {C.}~\bibnamefont {Grebing}}, \bibinfo
		{author} {\bibfnamefont {B.}~\bibnamefont {Lipphardt}}, \bibinfo {author}
		{\bibfnamefont {S.}~\bibnamefont {Weyers}}, \bibinfo {author} {\bibfnamefont
			{V.}~\bibnamefont {Gerginov}}, \bibinfo {author} {\bibfnamefont
			{N.}~\bibnamefont {Huntemann}}, \bibinfo {author} {\bibfnamefont
			{C.}~\bibnamefont {Hagemann}}, \bibinfo {author} {\bibfnamefont
			{A.}~\bibnamefont {Al-Masoudi}}, \bibinfo {author} {\bibfnamefont
			{S.}~\bibnamefont {H\"{a}fner}}, \bibinfo {author} {\bibfnamefont
			{S.}~\bibnamefont {Vogt}}, \bibinfo {author} {\bibfnamefont {U.}~\bibnamefont
			{Sterr}},\ and\ \bibinfo {author} {\bibfnamefont {C.}~\bibnamefont
			{Lisdat}},\ }\href {https://doi.org/10.1088/1367-2630/16/7/073023} {\bibfield
		{journal} {\bibinfo  {journal} {New Journal of Physics}\ }\textbf {\bibinfo
			{volume} {16}},\ \bibinfo {pages} {073023} (\bibinfo {year}
		{2014})}\BibitemShut {NoStop}%
	\bibitem [{\citenamefont {Zhu}\ \emph {et~al.}(2025)\citenamefont {Zhu},
		\citenamefont {Zhai}, \citenamefont {Xie}, \citenamefont {Miao},
		\citenamefont {Yu}, \citenamefont {Kong}, \citenamefont {Song}, \citenamefont
		{Zhang}, \citenamefont {Yi}, \citenamefont {Cui}, \citenamefont {Jiang},
		\citenamefont {Yang}, \citenamefont {Jia}, \citenamefont {Yin}, \citenamefont
		{Liao}, \citenamefont {Shu}, \citenamefont {Peng}, \citenamefont {Xu},
		\citenamefont {Dai}, \citenamefont {Chen},\ and\ \citenamefont
		{Pan}}]{zhu2025}%
	\BibitemOpen
	\bibfield  {author} {\bibinfo {author} {\bibfnamefont {X.-Q.}\ \bibnamefont
			{Zhu}}, \bibinfo {author} {\bibfnamefont {X.}~\bibnamefont {Zhai}}, \bibinfo
		{author} {\bibfnamefont {Y.}~\bibnamefont {Xie}}, \bibinfo {author}
		{\bibfnamefont {Y.}~\bibnamefont {Miao}}, \bibinfo {author} {\bibfnamefont
			{H.-W.}\ \bibnamefont {Yu}}, \bibinfo {author} {\bibfnamefont {D.-Q.}\
			\bibnamefont {Kong}}, \bibinfo {author} {\bibfnamefont {W.-L.}\ \bibnamefont
			{Song}}, \bibinfo {author} {\bibfnamefont {Y.}~\bibnamefont {Zhang}},
		\bibinfo {author} {\bibfnamefont {H.}~\bibnamefont {Yi}}, \bibinfo {author}
		{\bibfnamefont {X.-Y.}\ \bibnamefont {Cui}}, \bibinfo {author} {\bibfnamefont
			{X.}~\bibnamefont {Jiang}}, \bibinfo {author} {\bibfnamefont
			{B.}~\bibnamefont {Yang}}, \bibinfo {author} {\bibfnamefont {J.}~\bibnamefont
			{Jia}}, \bibinfo {author} {\bibfnamefont {J.}~\bibnamefont {Yin}}, \bibinfo
		{author} {\bibfnamefont {S.}~\bibnamefont {Liao}}, \bibinfo {author}
		{\bibfnamefont {R.}~\bibnamefont {Shu}}, \bibinfo {author} {\bibfnamefont
			{C.-Z.}\ \bibnamefont {Peng}}, \bibinfo {author} {\bibfnamefont
			{P.}~\bibnamefont {Xu}}, \bibinfo {author} {\bibfnamefont {H.-N.}\
			\bibnamefont {Dai}}, \bibinfo {author} {\bibfnamefont {Y.-A.}\ \bibnamefont
			{Chen}},\ and\ \bibinfo {author} {\bibfnamefont {J.-W.}\ \bibnamefont
			{Pan}},\ }\bibfield  {journal} {\bibinfo  {journal} {Optica}\ }\href
	{https://doi.org/10.1364/OPTICA.568436} {10.1364/OPTICA.568436} (\bibinfo
	{year} {2025})\BibitemShut {NoStop}%
	\bibitem [{\citenamefont {Dick}(2337)}]{dick1989local}%
	\BibitemOpen
	\bibfield  {author} {\bibinfo {author} {\bibfnamefont {G.~J.}\ \bibnamefont
			{Dick}},\ }in\ \href
	{http://www.ion.org/publications/abstract.cfm?jp=p&articleID=15462} {\emph
		{\bibinfo {booktitle} {Proceedings of the 19th {{Annual Precise Time}} and
				{{Time Interval Systems}} and {{Applications Meeting}}}}}\ (\bibinfo {year}
	{1989/12/337})\ pp.\ \bibinfo {pages} {133--147}\BibitemShut {NoStop}%
\end{thebibliography}
\end{document}